\documentclass[preprint,aps,prb,twocolumn,amsmath,showpacs,10pt,superscriptaddress,onlinecite]{revtex4}
\pdfoutput=1
\usepackage{graphicx}
\usepackage{hyperref}

\begin{document}
\preprint{Original publication available at prb.aps.org: Phys. Rev. B 90,
035121 (2014); DOI:10.1103/PhysRevB.90.035121}

\title{\textit{Ab initio} model of optical properties of two-temperature warm dense matter}

\author{B. Holst}
\affiliation{
CEA, DAM, DIF, 91297 Arpajon, France
}
\affiliation{
LULI \'{E}cole Polytechnique, CNRS, CEA, UPMC, 91128 Palaiseau, France 
}

\author{V. Recoules}
\affiliation{
CEA, DAM, DIF, 91297 Arpajon, France
}
\affiliation{
Luth, Observatoire de Paris, 92195 Meudon, France
}
\author{S. Mazevet}
\affiliation{
Luth, Observatoire de Paris, 92195 Meudon, France
}
\author{M. Torrent}
\affiliation{
CEA, DAM, DIF, 91297 Arpajon, France
}
\author{A. Ng}
\affiliation{
Department of Physics and Astronomy, University of British Columbia, Vancouver, British Columbia, Canada
}

\author{Z. Chen}
\affiliation{
Department of Electrical and Computer Engineering, University of Alberta, Edmonton, Alberta, Canada
}
\author{S. E. Kirkwood}
\affiliation{
Department of Physics, University of Ottawa, Ottawa, Ontario, Canada
}
\author{V. Sametoglu}
\affiliation{
Department of Electrical and Computer Engineering, University of Alberta, Edmonton, Alberta, Canada
}
\author{M. Reid}
\affiliation{
Department of Physics, University of Northern British Columbia, Prince George, British Columbia, Canada
}
\author{Y.~Y.~Tsui}
\affiliation{
Department of Electrical and Computer Engineering, University of Alberta, Edmonton, Alberta, Canada
}
\date{\today}

\begin{abstract}
We present a model to describe thermophysical and optical properties of two-temperature systems consisted of heated electrons and cold ions in a solid lattice  that occur during ultra-fast heating experiments. Our model is based on \textit{ab initio} simulations within the framework of density functional theory. The optical properties are obtained by evaluating the Kubo-Greenwood formula. By applying the material parameters of our \textit{ab initio} model to a two temperature model we are able to describe the temperature relaxation process of a femtosecond-laser heated gold and its optical properties within the same theoretical framework. Recent time-resolved measurements of optical properties of ultra-fast heated gold revealed the dynamics of the interaction between femtosecond laser pulses and solid state matter.
Different scenarios obtained from simulations of our study are compared with experimental data.~\cite{Chen2013} 
\end{abstract}


\maketitle

\section{Introduction}

The temporal behavior of non-equilibrium states of metals in the Warm Dense Matter (WDM) regime is under intense investigations using advanced experimental pump-probe techniques.~\cite{Sokolowski-Tinten2003, Uteza2004, Widmann2004,  Ping2006, Ao2006, Kandyla2007, Ping2008, Ernstorfer2009, Ping2010, Cho2011, Chen2013} In these experiments WDM is created with the irradiation of metals by ultra-short laser pulses. While the electrons  are heated during laser absorption in typical durations of 40-150~fs, the ions remain cold for a much longer period of time because the process of electron heating by the laser pulse is much faster than the kinetic energy transfer to the ionic degrees of freedom by means of electron-phonon coupling.~\cite{Fann1992, Fann1992a, Hohlfeld2000, Bonn2000, Hohlfeld1997, Suarez1995, Juhasz1993} As a consequence, a transient state of non-equilibrium matter consisting of these two subsystems with different temperatures is created by this procedure. This state can be observed prior to the disassembly of the metal by hydrodynamic expansion. During this time the excited electrons are expected to thermalize within one picosecond~\cite{Fann1992, Fann1992a} to form an equilibrium Fermi distribution with a well-defined temperature. Simultaneously, heat transfer from the electrons to the ions will occur. This can be described by rate equations in the well-known two-temperature model (TTM) that is widely used to describe laser-heated materials.~\cite{Anisimov1974a, Mazevet2005} The heat transfer rate can be accounted for by electron-phonon scattering where electron kinetic energy is transferred to the degrees of freedom of bound ions, described as phonons.~\cite{Allen1987, Wang1994, Lin2008} Other approaches consider the scattering of electrons with plasmons~\cite{Dharma-wardana1998, Vorberger2012} that is applicable at higher ion temperatures.

An important quantity for characterizing the WDM state is electrical conductivity. It has been calculated for plasmas at high temperatures using the Spitzer theory.~\cite{Spitzer1953} In the low-temperature, high-density regime characteristic of WDM, Spitzer theory becomes invalid and a many-particle theory is necessary. Lee and More~\cite{Lee1984} used the Boltzmann equation in relaxation time approximation. Another approach was the extended Ziman formula,~\cite{Ziman1961} which was successfully used to obtain the electrical conductivity of warm dense metals.~\cite{Rinker1985, Rinker1988} The combination of the Ziman formula and density functional theory has enabled the treatment of electron-ion plasmas over a wide range of thermodynamic parameters.~\cite{Perrot1987, Perrot1995, Dharma-wardana2006} Measurements of the DC conductivity in aluminum and copper plasmas at a few thousand Kelvin~\cite{DeSilva1998} have revealed significant deviations from the Lee-More model. The conductivity in  that particular region of the phase diagram is governed by the metal-nonmetal transition, as shown by Desjarlais \textit{et al.}~\cite{Desjarlais2002} In their model they applied density-functional-theory molecular dynamics simulations (DFT-MD) in combination with the Kubo-Greenwood formula to describe the behavior near that phase transition.  This method was used earlier to treat semiconductors~\cite{Vast1995, Holender1995} and metals.~\cite{Silvestrelli1999}

Since then DFT-MD simulations in combination with the Kubo-Greenwood formula has proven to be an effective tool for calculating electrical conductivity of WDM. In particular, the ability to obtain frequency dependent ac conductivity turned out to be useful when optical properties, i.e. the complex ac conductivity,  of WDM states produced by isochoric laser heating in an ultrathin gold foil were measured.~\cite{Widmann2004} In their simulations of non-equilibrium conditions pertinent to the experiment, Mazevet \textit{et al.}~\cite{Mazevet2005} showed that the calculated  ac conductivity value agrees well with measurements if the simulated state was taken to be a combination of heated electrons and a cold ion lattice. Such a state was also observed experimentally by Ping \textit{et al.}~\cite{Ping2006} Despite their importance, both experiments did not provide sufficiently precise characterization of the temporal evolution in optical properties. Only recently the time-dependent non-equilibrium optical properties of gold were obtained in single-shot pump-probe measurements for a duration of picoseconds and with a temporal resolution of 540~fs.~\cite{Chen2013} 

In this paper we present an \textit{ab initio} model based on DFT to calculate material parameters like the electron heat capacity and the electron-ion coupling factor on the one hand and the optical conductivity on the other hand within the same theoretical framework. The model is well suited to be applied to non-equilibrium warm dense matter, where the term ``non-equilibrium'' refers to a state where electrons and ions are at different temperature. In our simulations we considered a system electron states in equilibrium. The ions remain in an fcc lattice structure while moving around their equilibrium position. Coupled with a two-temperature model, it has enabled us to model the temporal evolution of AC conductivity during thermal equilibration between electrons and ions in femtosecond-laser heated gold.~\cite{Chen2013}   

The paper will begin with a description of the theoretical model in section \ref{sec:model}.
  This is followed by the comparison of model results with recent experimental data~\cite{Chen2013} in section   \ref{sec:comparison} and conclusions in section \ref{sec:conclusion}.

The comparison of data from this experiment with predictions from simulations indicate, that the electron-ion coupling is much weaker than predicted by Lin \textit{et al.}~\cite{Lin2008} Measurements for graphite~\cite{White2012} point to  a similar  conclusion. 

\section{Theoretical Model}
\label{sec:model}

In this section we present the theoretical framework of our study. This includes electron structure calculations and calculations of the AC conductivity. Knowledge of the electronic structure allows us to obtain the electron heat capacity and the electron-ion coupling factor, which can be used to describe energy relaxation processes of systems with different electron and ion temperatures. This enables us to compare our model calculations for the AC conductivity with findings from dynamic experiments, see Sec. \ref{sec:comparison} . 
 
 \subsection{Electron structure calculations} 

The electron system is described using the \textit{ab initio} plane wave code \textsc{Abinit}~\cite{Gonze2009}, which can readily be implemented on high performance computers.~\cite{Bottin2008} Here, the quantum many-body Schr\"odinger equation is solved using the density functional theory (DFT) that reduces to an effective one-particle problem for the electron density. The main approximation needed is the formulation of the exchange-correlation function, for which we apply the local density approximation (LDA), that gives a lattice constant in best agreement with known data compared to other approximations.~\cite{Becke1988, Perdew1992} In order to increase computational efficiency the projector augmented wave method (PAW)~\cite{Torrent2008} was applied, using a PAW data set that has been extensively tested.~\cite{Dewaele2008} The used PAW data set considers 11 valence electrons per atom. Depending on the electron temperature up to 48 DFT bands were used for single-atom calculations, that were carried out to obtain the electron density of states (DOS), the heat capacity, and the properties of the phonons. To obtain those properties highly converged, a cutoff energy of up to 40~Ha (1090~eV) was used for the electron structure calculations.  To ensure converged results the Brillouin zone was sampled using a Monkhorst-Pack $\mathbf{k}$-point sampling~\cite{Monkhorst1976} of up to $32\times32\times32$.

To incorporate effects of the ion system at elevated temperatures, molecular dynamics (MD) simulation is also carried out. All simulation results presented in the present work were obtained using 108 atoms in a cubic simulation box with periodic boundary conditions. For two-temperature MD simulations the ion and electron temperatures were controlled independently.  While the electron temperature was fixed for every timestep in the simulation, the ion temperature was controlled by a Nos\'e-Hoover thermostat.~\cite{Nose1984}   Before reaching a stable ion temperature we had to allow the  system  to relax  for a few hundreds of timesteps.  After convergence of the ion structure has been reached, the ion temperature fluctuates around a well defined temperature and thermodynamic parameters as well as the conductivity can be obtained. Depending on the electron temperature we had to choose the number of DFT bands as high as 2048. We found a cutoff energy of 20~Ha (545~eV) to be sufficient for all MD simulations within the present work. All MD simulations were carried out at the $\Gamma$-point of the Brillouin zone.  

\begin{figure}[htb]
\includegraphics[width=\columnwidth]{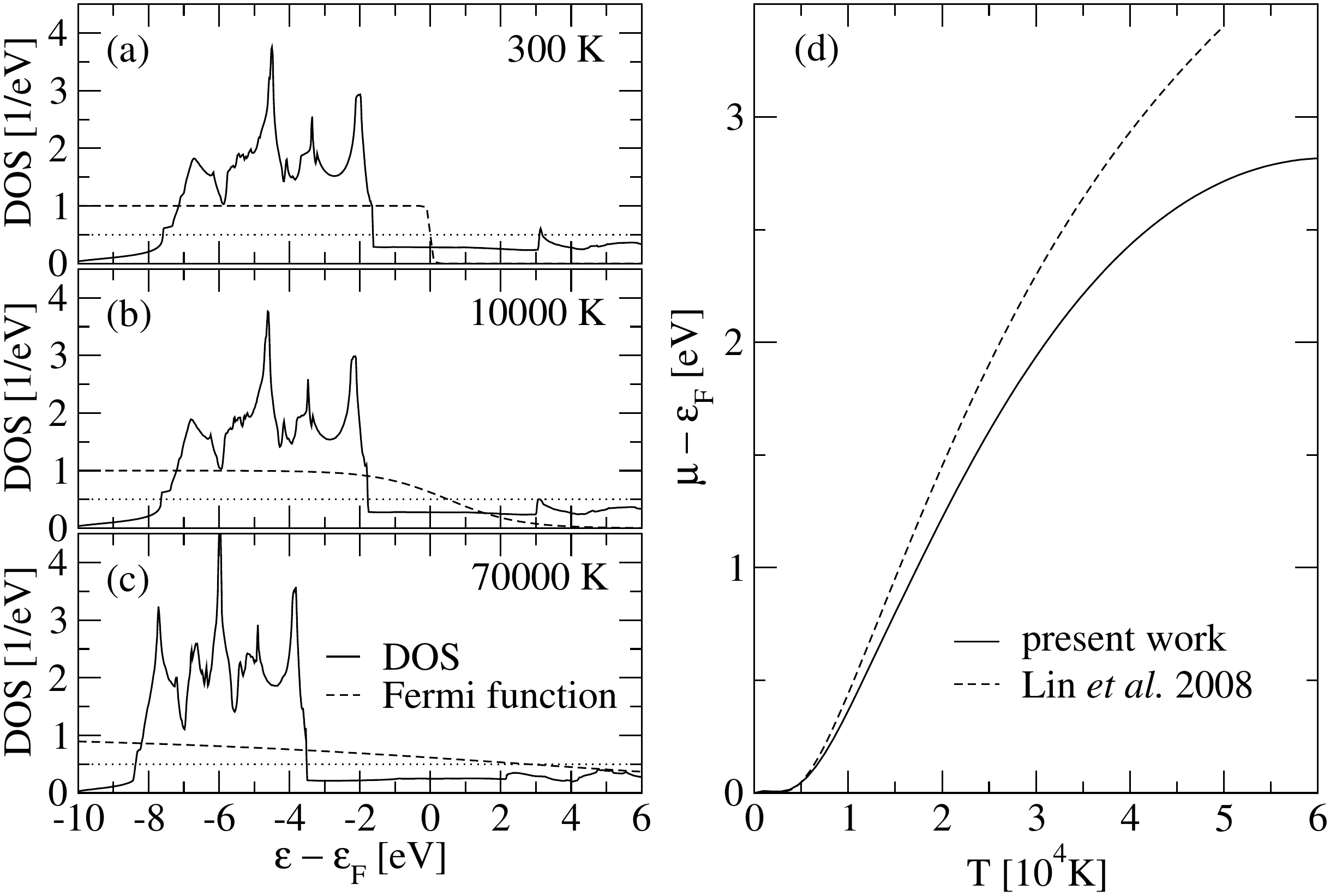}
 \caption{ (a) - (c) Density of states (solid) of fcc gold for several electron temperatures and solid density. Also shown is the Fermi distribution function (dashed). (d) Chemical potential of gold computed with \textsc{Abinit} (solid) taking into account the change of the DOS with $T_e$ compared to the result of Lin~\textit{et~al.}~\cite{Lin2008} (dashed).}
\label{fig:DOSmu}
\end{figure}

The DOS for three different electron temperatures is shown in Figs. \ref{fig:DOSmu}~(a-c). Here, the origin of the energy scale was chosen to be the Fermi energy which is the chemical potential at 0~K. Also shown is the Fermi distribution with a width determined by the electron temperature. The chemical potential,  which is obtained by the conservation of the electron number $N = \int g(\varepsilon,T) f(\varepsilon,\mu,T)\label{eq:partncons}$, changes with temperature and is shown in Fig.~\ref{fig:DOSmu}(d). 
At the lowest temperature shown, the DOS in the vicinity of the Fermi energy is  slowly increasing depicting a behavior similar to that of a free electron gas, which itself would be represented by a square-root function. This region of the DOS is populated by the 6s electrons of the gold atom. For physical properties governed only by contributions arising from a region of the DOS where the temperature-derivative of the Fermi function is non-zero, the system behaves like a free electron gas.
At higher temperatures the region with changing Fermi function is broadened and overlaps with the DOS structure at lower energy that is dominated by the 5d electrons.  The system will then behave differently than a free electron gas, as will be seen later.

It should also be noted that the DOS is shifted towards lower energies as temperature increases. 
Furthermore, due to the increasing width of the Fermi function, the chemical potential has to be increased in order to satisfy the conservation of particle number. This is compensated by the shift of the DOS towards lower energies, which causes the chemical potential to increase further at high temperature. Such a behavior is not seen in the simulations of Ref.~\onlinecite{Lin2008} where a constant DOS is assumed.

\subsection{Electron heat capacity}

To compute the electron heat capacity we first perform electron structure calculations for a single atom in a simulation box with periodic boundary conditions, which results in a perfect face centered cubic lattice.     The DFT minimization scheme with respect to the free energy of the electrons yields directly the internal energy.  By carrying out DFT calculations for different electronic temperatures we then obtain the internal energy as a function of temperature, which allows us to calculate the electron heat capacity from its temperature derivative using Eq.~(\ref{eq:ce}), 
\begin{equation}
 C_e=\left(\frac{\partial U}{\partial T_e}\right)_V \label{eq:ce}
\end{equation}

This expression contains all thermal effects within DFT, including the temperature dependence of the density of states.  

\begin{figure}[htb]
\includegraphics[width=0.7\columnwidth]{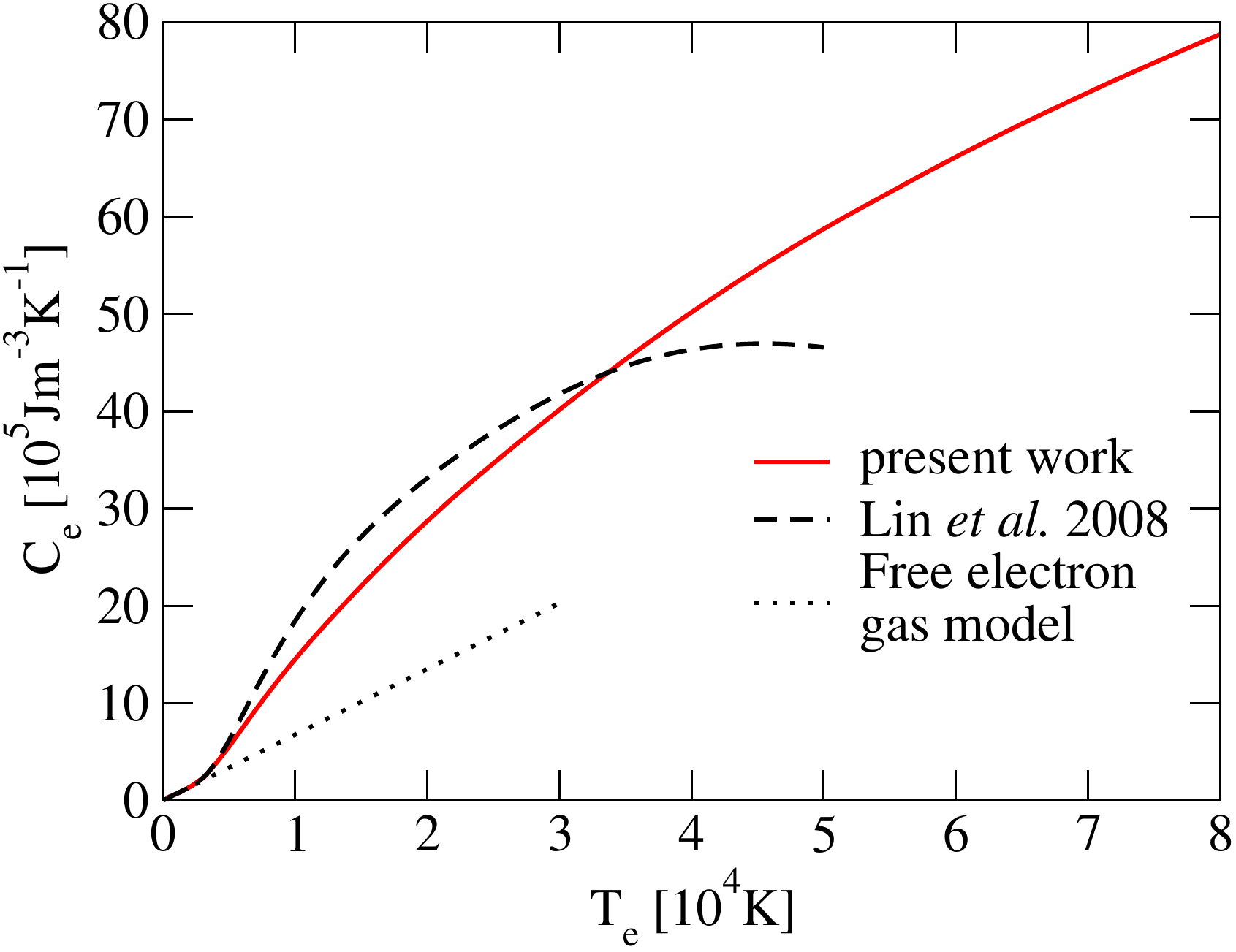}

\caption{Electron heat capacity of gold as a  function of electron temperature. The results of the present work are shown as solid red curve. Also shown are the data of Lin~\textit{et al.}~\cite{Lin2008} (dashed) and the free electron gas approach with $\gamma_e$=67.6~J/m$^{3}$K$^{2}$ (dotted).}
\label{fig:cve}
\end{figure}

Our results  are presented in Fig.~\ref{fig:cve}. At low electron temperature the curve of $C_e$ is linear with a slope very similar to that found in experiment.~\cite{Chen2013}  This is the same behavior as that expected for a free electron gas.  The explanation for this is that the region with non-zero derivative of the Fermi function is very small and located where the DOS shows a similar shape  as that of the free electron gas,  originating from the  6s electrons.   As the temperature increases the curve of $C_e$   becomes nonlinear.  In this situation the derivative of the Fermi function is non-zero in the same region where the DOS shows the largely structured feature due to 5d electrons.  This causes a significant deviation from the free electron gas behavior.  $C_e$ is now increasing faster with temperature. A similar behavior is also shown by the results of Ref.~\onlinecite{Lin2008} except for a local maximum at $\approx4\times 10^4$~K.  This difference can be traced to the use of a temperature-independent DOS.

\begin{figure}[htb]
\includegraphics[width=0.5\columnwidth]{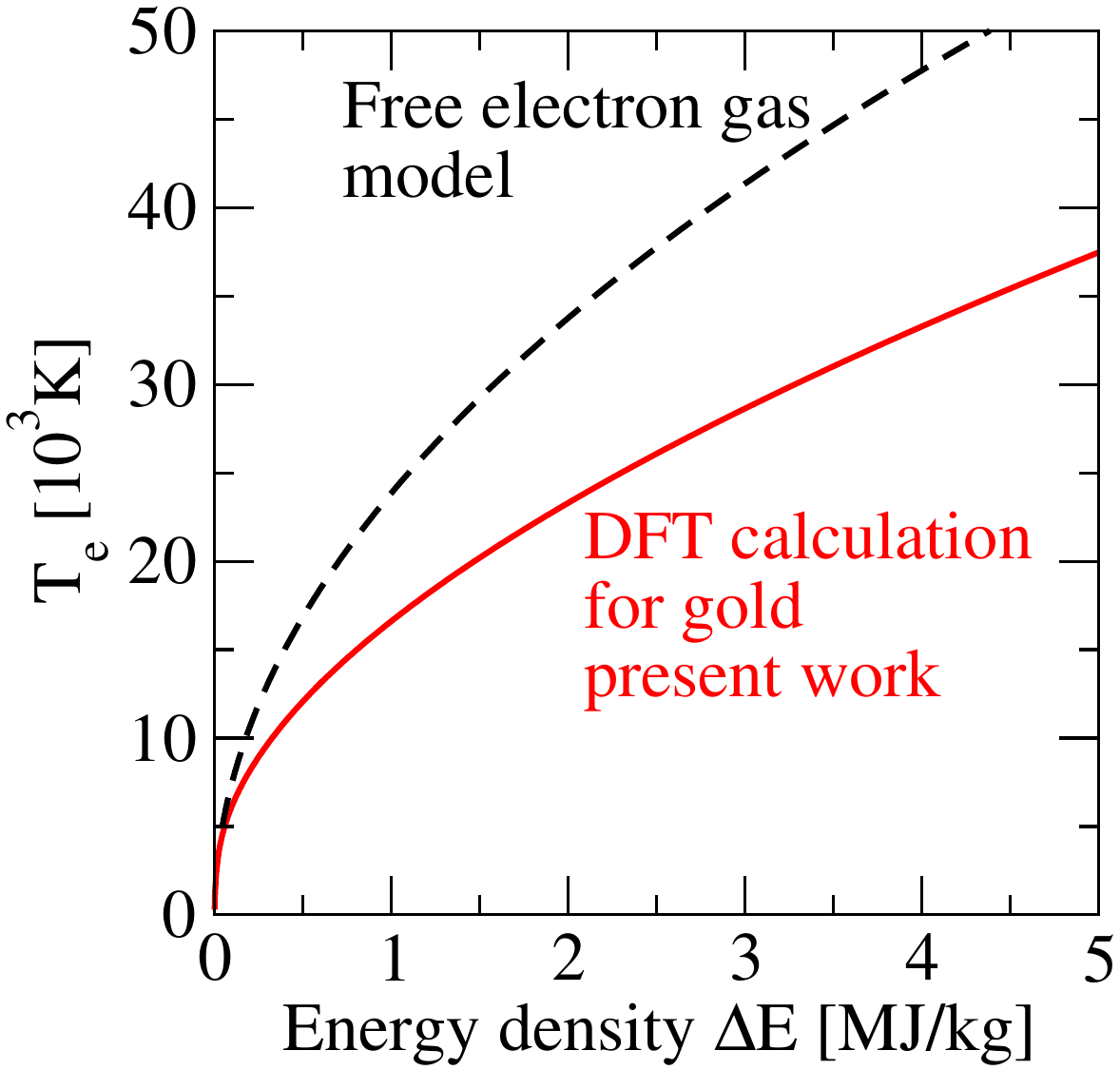}
 \caption{Electron temperature as a function of absorbed energy. The different curves were calculated using different models: FEG model for $C_e$ (dashed black line) and $C_e$ using \textsc{Abinit} (solid red line). }
\label{fig:T-initial}
\end{figure}

The knowledge of the electron heat capacity enables us to calculate the excited electron temperature as a function of absorbed energy density.  It is worth to examine this quantity, because it can be taken to be the initial temperature of the electrons in experiments after the pump pulse is absorbed and before significant temperature relaxation has occurred.  At such a time, we assume that all of the absorbed laser energy resides in the electron system.  This assumption tends to overestimate $T_e$ since the temperature relaxation process would have started during the time of the pump laser pulse.  However, the discrepancy is expected to be small.  The initial electron temperature thus calculated is shown as a function of absorbed laser energy density is shown in Fig.~\ref{fig:T-initial}.  Also displayed are the results using $C_e$ from the free electron gas model (FEG).  As expected, the lower $C_e$ values from the FEG model also leads to much higher predicted electron temperatures. 

\subsection{Electron-ion coupling factor}
\label{sec:gei}

Following Allen~\cite{Allen1987} and Wang \textit{et al.}~\cite{Wang1994} the process of electron-ion coupling is treated via an energy transfer rate from electrons to phonons. This is assumed to be proportional to the temperature difference between the lattice and electrons. The electron-phonon coupling factor $G_{ei}$ is introduced as follows: 
\begin{equation}
 G_{ei}\left(T_l-T_e\right)=\left. \frac{\partial E}{\partial t}\right|_{ep} \mathrm{.}
\end{equation}

Here we use the definition of the heat transfer rate by Allen~\cite{Allen1987}:

\begin {equation}
 \left. \frac{\partial E}{\partial t}\right|_{ep}=\frac{4\pi}{\hbar}
\sum_{k,k'}\hbar\omega_Q\left|M_{kk'}\right|^2S(k,k')\delta(\varepsilon_k-\varepsilon_{k'}+\hbar\omega_Q) \label{eq:e-transfer}
\end {equation}

where $k$ and $Q$ are electron and phonon quantum numbers and the matrix element $M_{k,k'}$ describes the scattering probability of an initial electron state $k$ at energy $\varepsilon_k$ to a state $k'$ at energy $\varepsilon_{k'}$. The factor $S(k,k')=(f_k-f_{k'})n_Q-f_{k'}(1-f_k)$ contains the electron and phonon occupation numbers represented by the Fermi distribution function $f_k=1/\left\lbrace\exp\left[(\varepsilon_k-\mu)/k_B T_e\right]+1\right\rbrace$ and the Bose distribution function $n_Q=1/\left[\exp\left(\omega_Q/k_B T_l\right)-1\right]$ respectively.

Using the definition of the Eliashberg spectral function 
\begin{eqnarray}
 \alpha^2F(\varepsilon',\varepsilon, \Omega)&=&\frac{2}{\hbar g(\varepsilon_\text{F})}
\sum_{k,k'}\left|M_{kk'}\right|^2\\
&&\times\delta(\omega_Q-\Omega)\delta(\varepsilon_k-\varepsilon)\delta(\varepsilon_{k'}-\varepsilon')\nonumber
\end{eqnarray}
in combination with Eq.~(\ref{eq:e-transfer}) we get  by multiplying  with the three integrals $\int\mathrm{d}\varepsilon\delta(\varepsilon_k-\varepsilon)$, $\int\mathrm{d}\varepsilon\delta(\varepsilon_{k'}-\varepsilon')$ and $\int\mathrm{d}\Omega\delta(\omega_Q-\Omega)$  the following expression:
\begin{eqnarray}
 G_{ei}&=&\frac{2\pi g(\varepsilon_\text{F})}{T_l-T_e}\int_0^\infty\mathrm{d}\Omega\hbar\Omega
\int_{-\infty}^\infty\mathrm{d}\varepsilon\int_{-\infty}^\infty\mathrm{d}\varepsilon' \alpha^2F(\varepsilon,\varepsilon',\Omega) \nonumber\\
&&\times S(\varepsilon,\varepsilon')\delta(\varepsilon-\varepsilon'+\hbar\Omega) \label{eq:gei-full} \text{.} 
\end{eqnarray}
Together with the energy conservation $\varepsilon'=\varepsilon+\hbar\Omega$ the thermal factor is now 
\begin{equation}
S(\varepsilon,\varepsilon')=[f(\varepsilon)-f(\varepsilon')][n(\varepsilon'-\varepsilon,T_l)-n(\varepsilon'-\varepsilon,T_e)] \mathrm{.}
\end{equation}

We employ the approximation suggested by Wang \textit{et al.}~\cite{Wang1994} for the electron-phonon spectral function, $a^2F(\varepsilon,\varepsilon+\hbar\Omega,\Omega)= \left[g(\varepsilon)g(\varepsilon+\hbar\Omega)/g^2(\varepsilon_\text{F})\right] \alpha^2F(\varepsilon_\text{F},\varepsilon_\text{F},\Omega)$ where the last function is the electron-phonon spectral function at the Fermi energy which can by obtained in the framework of density functional perturbation theory. Using this approximation one of the energy integrals can be solved and Eq.~(\ref{eq:gei-full}) becomes
\begin{eqnarray}
\label{eq:gei-wang}
G_{ei}&=&\frac{2\pi\hbar}{T_l-T_e} \int_o^\infty \mathrm{d}\Omega\alpha^2F(\Omega)\Omega
\int_{-\infty}^{\infty}\textrm{d}\varepsilon \frac{ g(\varepsilon)g(\varepsilon+\hbar\Omega)}{g(\varepsilon_F)} \nonumber\\
&&\times  S(\varepsilon,\varepsilon+\hbar\Omega) \mathrm{.} \label{eq:gei-a2f}
\end{eqnarray}

Eq.~(\ref{eq:gei-wang}) can be furthermore simplyfied by means of the high temperature expansion of the Bose functions, i.e. $k_B T \gg \hbar\Omega$. This allows to rewrite the thermal factor which becomes
\begin{equation}
S(\varepsilon,\varepsilon+\hbar\Omega)=\frac{k_B}{\hbar\Omega}(T_l-T_e)\left(f(\varepsilon)-f(\varepsilon+\hbar\Omega)\right) \mathrm{.}
\end{equation}
Introducing the second moment of the of $\alpha^2F(\Omega)$, which is $\lambda\langle\omega^2\rangle = 2\int_0^\infty\mathrm{d}\Omega \alpha^2F(\Omega)\Omega$, assuming the energy range of the phonon frequencies to be much smaller than considered electron energies, i.e. $g(\varepsilon)\approx g(\varepsilon+\hbar\Omega)$ and $\partial f/\partial\varepsilon=[f(\varepsilon+\hbar\Omega)-f(\varepsilon)]/\hbar\Omega$, we can write down the electron-phonon coupling factor in the version used by Lin \textit{et al.}~\cite{Lin2008}:  

\begin{equation}
G_{ei}=\frac{\pi\hbar k_B\lambda\langle\omega^2\rangle}{g(\varepsilon_F)}
\int_{-\infty}^{\infty}g^2(\varepsilon,T)\left(-\frac{\partial f}{\partial\varepsilon}\right)\textrm{d}\varepsilon \mathrm{.} \label{eq:gei-lin}
\end{equation}

\begin{figure}[htb]
 \includegraphics[width=0.7\columnwidth]{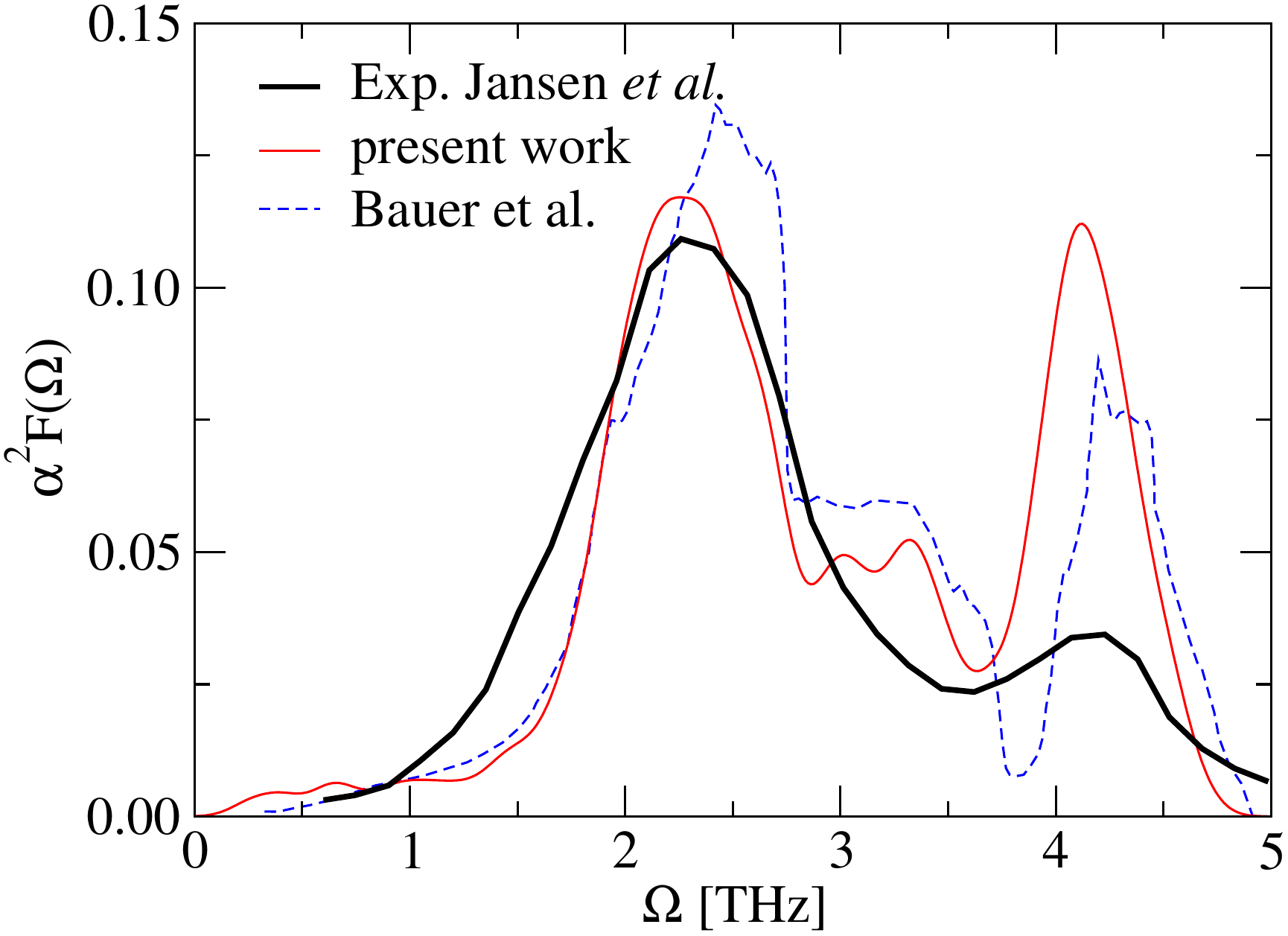}
  \caption{Electron-phonon spectral function of fcc gold at 300K obtained in experiment~\cite{Jansen1977} (bold) compared to linear response DFT results from this work (red solid) and Bauer \textit{et al.}~\cite{Bauer1998} (blue dashed).}
 \label{fig:a2f}
 \end{figure}

The disadvantage of using Eq.~(\ref{eq:gei-a2f}) instead of Eq.~(\ref{eq:gei-lin}) is the necessity of knowing the complete electron-phonon spectral function. This can be avoided by the latter formulation. Instead the electron-ion coupling is expressed by its second moment, which can be either obtained from experiments or approximated by the Debye temperature.~\cite{Lin2008} This approximation may introduce additional errors, which have been found small for aluminum and other metals at $T_e = 300$~K and becomes negligible at $T_e = 1000$~K.~\cite{Lin2008} Equally to the findings of Lin \textit{et al.}~\cite{Lin2008} for aluminum we find no significant changes caused by this approximations for gold. We find in addition that for gold the difference is negligible even at room temperature. To analyze the effect of the $\Omega$-integration the range of phonon frequencies has to be compared to the energy range of the electronic DOS. For gold we found that the electron-phonon spectral function is non-zero only below 5~THz, that is 20~meV (see Fig.~\ref{fig:a2f}), which is rather small compared to eigenenergies of the electronic states. Within this value the electronic DOS can be assumed almost constant, which is why the $\Omega$-integration can be reduced to a constant, which leads to Eq.~(\ref{eq:gei-lin}).  

In their previous work, Lin \textit{et al.}~\cite{Lin2008} found values of $G_{ei}$ significantly higher than indicated by experimental findings.~\cite{Hohlfeld2000} This would reduce the temperature relaxation time drastically. We calculated $G_{ei}$ in a similar way, but included the temperature dependence of the electronic DOS into Eq.~(\ref{eq:gei-lin}), which was neglected by the authors of the former work. The difference between our results is caused by this temperature dependence, similar to the case of electron heat capacity.

The calculation of the electron-ion coupling factor $G_{ei}$ by Eq.~(\ref{eq:gei-a2f}) requires the knowledge of the electron-phonon spectral function $\alpha^2F(\Omega)$, which can be calculated within the framework of linear response density functional perturbation theory calculations that are implemented in \textsc{Abinit}.~\cite{Gonze1997, Gonze1997a} The algorithm for the electron-phonon interaction is based on an analysis of the phonon line widths. These are computed via matrix elements for the transition between electron Bloch states whose energy differs by the phonon frequency (see Ref~\onlinecite{Savrasov1996, Liu1996} for details of the method). We performed our calculations for a single atom in a simulation box with periodic boundary conditions in order to simulate a perfect fcc lattice. The electronic wavefunctions were sampled on a $16\times16\times16$  $\mathbf{k}$-point grid and represented by a plane-wave expansion with a cutoff of 40 Ha. A norm-conserving pseudopotential~\cite{Troullier1991} for the electron-ion interaction and the local-density approximation (LDA) to account for the exchange-correlation functional was used. The perturbations of the atoms were resolved on a $8\times8\times8$ $\mathbf{q}$-point grid. 

The calculated electron-phonon spectral function $\alpha^2F(\Omega)$ is plotted in Fig.~\ref{fig:a2f}. There are 2 pronounced phonon frequencies at about 2.5~THz and 4~Thz. The current ABINIT results reproduce the observed frequencies of these features~\cite{Johnson1972}, while the results of Bauer \textit{et al.}~\cite{Bauer1998} appear slightly shifted. The intensity of the lower frequency feature is well reproduced by the ABINIT result, but the intensity at the higher frequency is overestimated. The second moment obtained from the ABINIT results is $\lambda\langle\omega^2\rangle=21\pm2$ meV$^2$ which is in agreement with the previously measured value~\cite{Brorson1987} of $23\pm4$ meV$^2$.

 \begin{figure}[htb]
 \includegraphics[width=0.7\columnwidth]{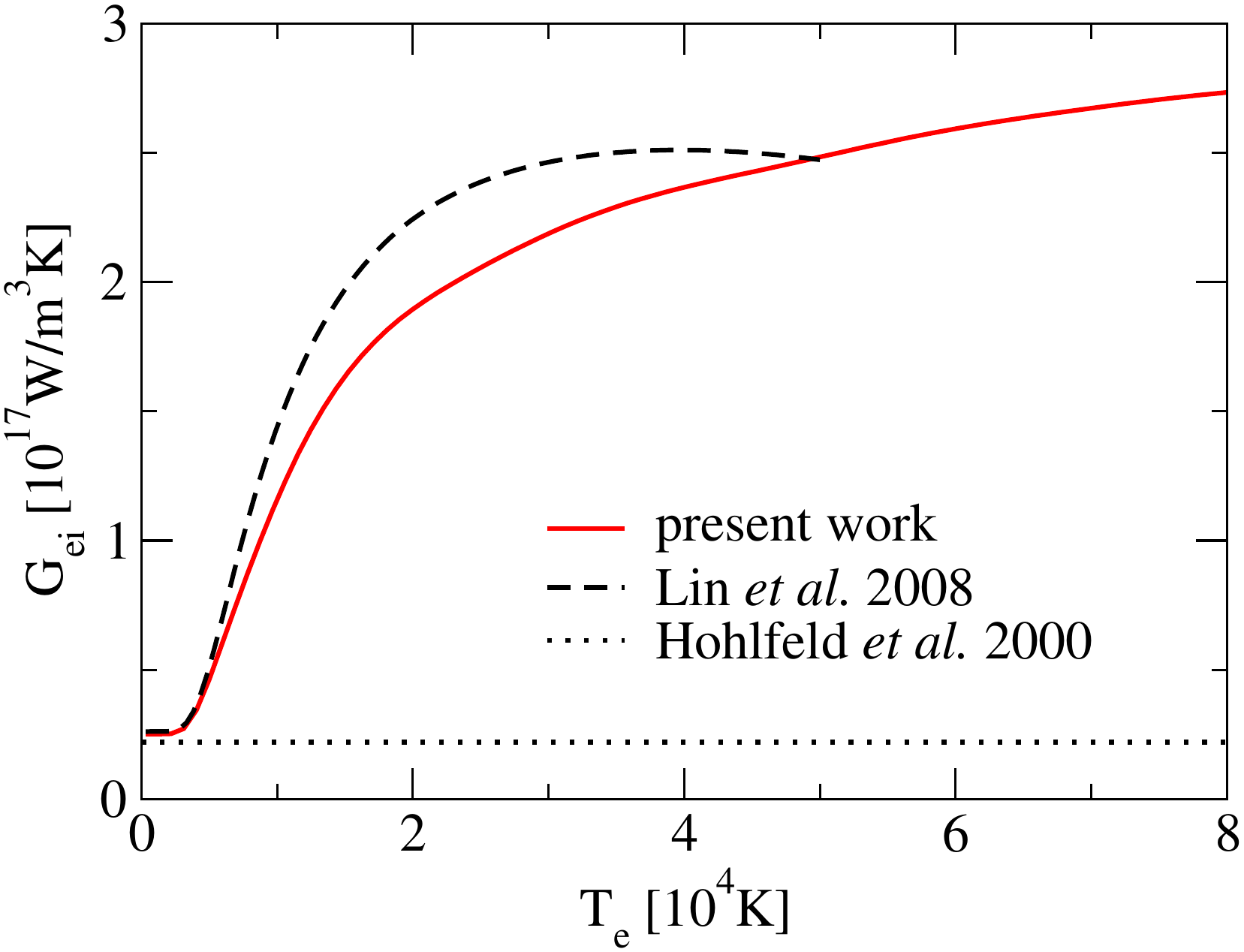}
  \caption{Electron-phonon coupling factor of fcc gold as a function of electron temperature at a lattice temperature of 300 K. Results from this work (red solid) are compared to findings of Lin \textit{et al.}~\cite{Lin2008} and to the constant factor found in experiments at lower energy density~\cite{Hohlfeld2000} (dotted) }
 \label{fig:gei}
 \end{figure}

In Fig.~\ref{fig:gei} our results for the electron-phonon coupling factor are plotted. 
At low temperatures up to 2000~K $G_{ei}$ is found to be constant, which is in agreement with both the previous theoretical prediction and experimental values. At higher temperatures the curves show a sharp rise to values ten times higher than the low-temperature value. While the curve of the current work shows a continuous increase until $80,000$~K, a maximum at about $40,000$~K can be seen in the curve of Lin~\textit{et~al.} This difference is again caused by 
our inclusion of temperature dependence in our electronic  DOS. It is worth to mention, that if we keep the DOS constant we could reproduce the results of Lin \textit{et al.} With its much higher $G_{ei}$ values, our current model would yield a much shorter equilibration time between electrons and ion.  This has significant impact on the temporal evolution of electron and ion temperatures, particularly at high values of absorbed laser energy density.

\subsection{AC conductivity} 

From MD simulations with 108 atoms at stable ion temperature we took several snapshots of the ion positions. These were used to start precise static DFT calculations in order to apply the Kubo-Greenwood formula.~\cite{Mazevet2010, Holst2011} The results from different snapshots from the same simulation run were averaged in order to reduce statistical uncertainties. 

\begin{eqnarray}\label{eq:kg}
\sigma_\text{r}(\omega) &=& \frac{2\pi e^2\hbar^2}{3 m^2 \omega V}
\sum_{\mathbf k} W({\mathbf k}) \sum_{j=1}^{N_\text{B}}\sum_{i=1}^{N_\text{B}}\sum_{\alpha=1}^3
\left[ F(\epsilon_{i,{\mathbf k}})-F(\epsilon_{j,{\mathbf k}})\right] \nonumber\\
&& \times |\langle\Psi_{j,{\mathbf k}}|\nabla_\alpha|\Psi_{i,{\mathbf k}}\rangle|^2
\delta(\epsilon_{j,{\mathbf k}}-\epsilon_{i,{\mathbf k}}-\hbar\omega) 
\end{eqnarray}

Here $e$ and $m$ are the electron charge and  mass. The summations over $i$ and $j$ run over $N_\text{B}$ discrete Kohn-Sham eigenstates considered in the electronic structure calculation. The three spatial directions are averaged by the $\alpha$ sum and $V$ is the volume of the simulation box. $F(\epsilon_{i,{\mathbf k}})$ is the Fermi distribution function and describes the occupation of the $i$th band corresponding to the energy $\epsilon_{i,{\mathbf k}}$ and the wave function $\Psi_{i,{\mathbf k}}$ at ${\bf k}$. The $\delta$-function has to be broadened because a discrete energy spectrum results from the finite simulation volume.~\cite{Desjarlais2002}

Integration over the Brillouin zone is performed by
sampling special ${\mathbf k}$-points,~\cite{Monkhorst1976} where $W({\mathbf k})$
is the respective weighting factor. A $4\times4\times4$ Monkhorst-Pack sampling grid was applied, i.e. the Brillouin zone was sampled at 32 ${\mathbf k}$-points. Simulating multiple atoms also introduces additional Kohn-Sham eigenstates per band, which allows for taking into account also intraband transitions, which are neglected in single atom conductivity calculations. As it was recently shown for sodium~\cite{Pozzo2011} an increase in the number of simulated atoms yields a convergence towards the correct value of DC conductivity which is usually underestimated for metals by DFT calculations. However, within the current work we had to limit our simulations to 108 atoms, resulting in a minimum distance of $\approx0.1$~eV between the Kohn-Sham eigenstates. As a consequence, the DC conductivity as well as AC conductivity at energies below 0.1~eV cannot be reproduced correctly. The AC conductivity at the experimental value of 1.55~eV is converged and agreement with experiments could be shown. 

The imaginary part $\sigma_\text{i}$ of the conductivity can be derived from the $\sigma_\text{r}$ using the Kramers-Kronig relation, i.e. the principal value of the integral over $\sigma_\text{r}$ is calculated.

\begin{equation}
\sigma_\text{i}(\omega)=-\frac{2}{\pi} \text{P}\int
\frac{\sigma_\text{r}(\nu)\omega}{(\nu^2-\omega^2)}d\nu 
\label{eq:kkr}
\end{equation}


\begin{figure}[htb]
\includegraphics[width=\columnwidth]{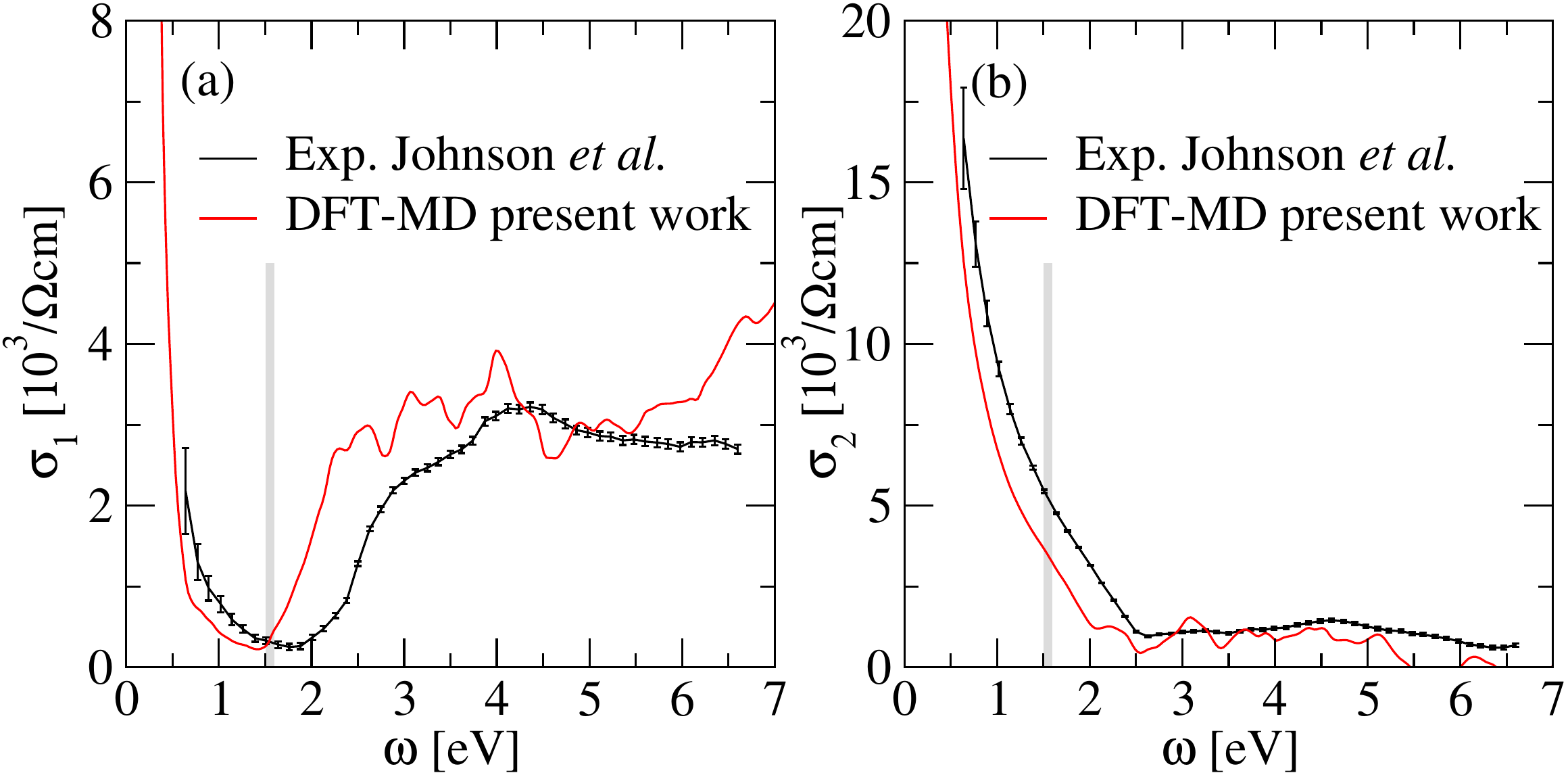}
 \caption{(a) Real and (b) imaginary part of the AC conductivity of gold at a photon energy of 1.55~eV and room temperature. The results of the current simulations aure compared to the measured data of Johnson \textit{et al.}~\cite{Johnson1972} }
\label{fig:sigma300K}
\end{figure}

In Fig.~\ref{fig:sigma300K} the  real and  imaginary  part  of the  gold AC conductivity at  ambient conditions  is depicted.   The experimental data are also included for a direct comparison.  It can be seen that the general characteristics of the measurement can be reproduced.  However, the DFT curves are clearly shifted towards the lower energies.  This is the result of the LDA approximation which is used for the exchange-correlation part  of the  energy  functional.  It leads to inaccuracies in the Kohn-Sham eigenvalues, producing a shift of the AC conductivity curve.   The influence of different approximations for the exchange-correlation potential was examined in another study on the band structure of gold.~\cite{Rangel2012} It was found that accuracies of the Kohn-Sham eigenvalues might be improved using hybrid HSE functional.  Since simulations with  the  HSE functional  are far more demanding computationally, we were not able to perform  those  calculations  for the  large number  of elevated  electron  temperatures needed  to describe the experiment.   Instead we rely on the LDA values to construct model predictions. At photon energy of 1.55~eV pertinent to the experiment of interest, the spectral shift has only minor impact on the real part of AC conductivity since the latter is almost constant in this region.   On the other hand, the impact is much larger on the imaginary part of AC conductivity because of its steep slope at 1.55~eV.  This renders its calculation less reliable.

Another problem in our model is the calculation of the imaginary part of AC conductivity from its real part using the Kramers-Kronig relation.  Such a procedure requires accurate values of the real part of AC conductivity down to the DC limit.   In our calculation based on 108 atoms in a unit cell, the DC limit is a factor of two below experimental value.  Test calculations with 256 atoms leads to no significant improvement.  It is well known that improvement with particle number in DFT calculations is not always guaranteed.  Results from Ref.~\onlinecite{Pozzo2011} have shown success with around 1000 atoms for some metals. Another study~\cite{Lambert2011} on high-density hydrogen plasma is finding even higher atom numbers are necessary to obtain converged results for the DC limit. Currently, calculations with such a number of atoms in a unit cell are not practical for gold.  On the other hand, the error in the calculation of the imaginary part of AC conductivity appears to be significant only at low temperature.
  
Both errors are expected to be smaller at high temperature, because on the one hand the influence of the approximation for the exchange-correlation functional becomes smaller~\cite{Faleev2006} and on the other hand it was shown that DFT can reproduce the DC conductivity for metals in the WDM regime well.~\cite{Desjarlais2002} We therefore think that our model will give reliable values for the prediction of the real part of the AC conductivity in WDM experiments. Although the imaginary part is expected to be less precise we will do a comparison with the measured values.

\section{Comparison to experiment for femtosecond-laser heated gold}
\label{sec:comparison}

\subsection{AC conductivity of initial heated states}
As the first application of our model, we calculate the AC conductivity of initial heated states in which all absorbed laser energy is assumed to reside in a thermalized electron subsystem as the ions remain at 300~K.  Such initial heated states are taken to occur 540~fs after the peak of the pump laser pulse in the experiment.~\cite{Chen2013}   The electron temperatures of these initial heated states are derived from the excitation energy density due to laser absorption as presented in Fig.~\ref{fig:T-initial}.

\begin{figure}[htb]
\includegraphics[width=\columnwidth]{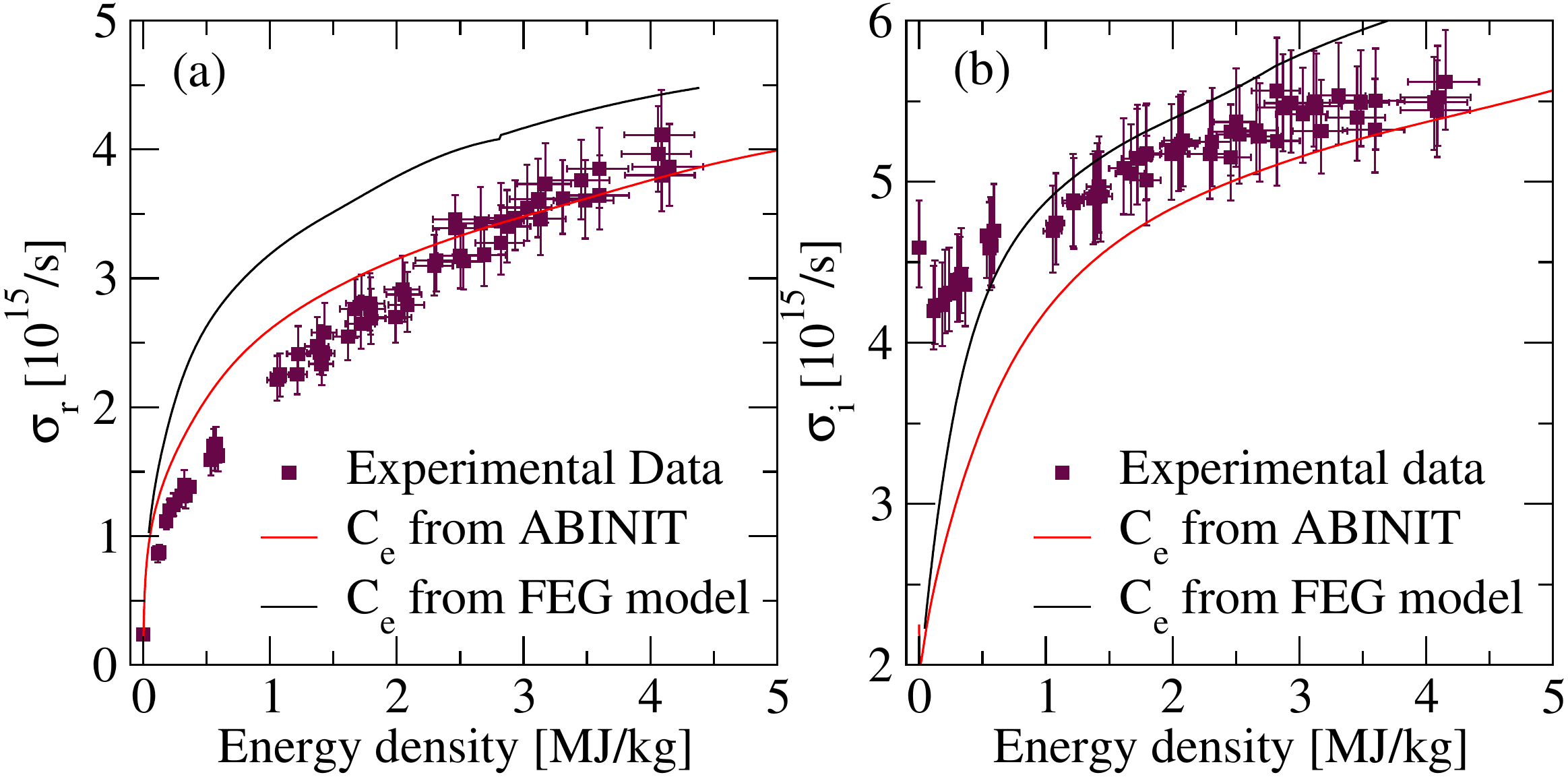}
 \caption{(a) Real and (b) imaginary part of the initial AC conductivity of gold at a photon energy of 1.55~eV as a function of excitation energy.}
\label{fig:initial-cond}
\end{figure}

In Fig.~\ref{fig:initial-cond} the comparison between measured data and model predictions for AC conductivity of the initial heated states is shown as a function of excitation energy density. There are two different theoretical curves: one using $C_e$ obtained from FEG model and the other using $C_e$ from \textsc{Abinit}. Looking at the real part in Fig.~\ref{fig:initial-cond}(a) it can be noted that the curve using the \textsc{Abinit} result is in better agreement with the experiment. Although there are some deviations around 1~MJ/kg the model is able to describe the measured data. It can be concluded by these results, that the presented model provides the real part of the AC conductivity in agreement with the measurements and that the \textsc{ABINIT} heat capacity is the more reasonable choice for the input to the TTM.

Looking at the imaginary part in  Fig.~\ref{fig:initial-cond}(b) a significant difference between experiment and model can be seen at excitation energy densities below $\sim 1$~MJ/kg. As discussed above, this difference results from the not correctly calculated value for the static limit of the real part. Nevertheless, also in this case the curve, which was obtained using the heat capacity derived from \textsc{Abinit} agreed better with the experimental data.

\subsection{AC conductivity during electron energy relaxation}
\label{subsec:relaxation}

The second application of our model is the calculation of AC conductivity during electron energy relaxation after the heated electrons have thermalized to a Fermi distribution with a well-defined electron temperature.  Here, the process of electron energy relaxation is described using a Two Temperature Model (TTM).~\cite{Anisimov1974a} Accordingly, the evolution of electron temperature $T_e$ and ion temperature $T_i$ are governed by two coupled equation:

  \begin{eqnarray*}
     C_e(T_e)\frac{\partial T_e}{\partial t} &=& 
-G(T_e)(T_e-T_i) + S(t)\\
     C_l(T_i)\frac{\partial T_i}{\partial t} &=& 
G(T_e)(T_e-T_i)
   \end{eqnarray*}

Electron thermal conduction is neglected since uniform heating of the ultrathin gold foil is produced by ballistic electron transport.~\cite{Chen2012}  The ion heat capacity $C_i$ is taken to be the lattice heat capacity at a constant value of $2.5\times 10^6$~J/m$^3$K, which is the Dulong-Petit limit of the Debye heat capacity for gold. The electron heat capacity $C_e$ and electron-ion coupling factor $G_{ei}$ are obtained from our model as describe above.   The initial ion temperature $T_i$ is 300~K and the initial electron temperature $T_e$ is determined by the absorbed laser energy density as shown in Fig.~\ref{fig:T-initial}.

\begin{figure}[htb]
\includegraphics[width=0.5\columnwidth]{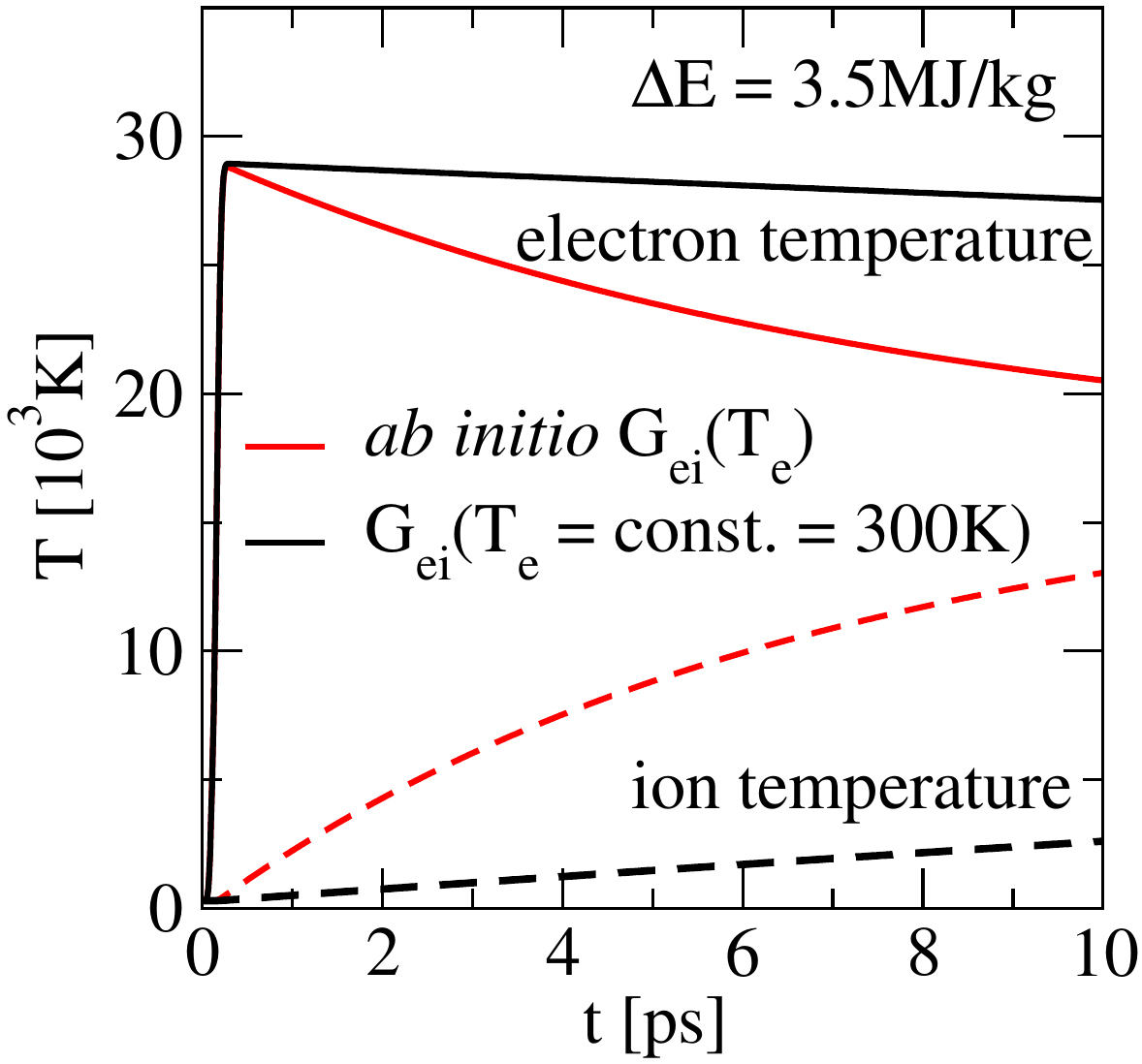}
 \caption{Temperature evolution in gold for an energy density of 3.5~MJ/kg and different input parameters for the TTM. The upper solid curves depict the electron temperature and the lower dashed the ion temperature. The different curves were calculated using a $G_{ei}$ from \textsc{Abinit} calculations (red) which is a function of $T_e$  and a constant $G_{ei}(T_e=300$~K$)$ (black) which does not depent on electron temperature. }
\label{fig:ttm}
\end{figure}

An example of our calculation is shown in Fig.~\ref{fig:ttm} for an excitation energy density of 3.5~MJ/kg.
The electrons are initially heated to approximately 30,000~K. At this temperature the temperature dependent $G_{ei}$ from our model is  significantly higher than the over $T_e$ constant one.  

This causes the relaxation process to be accelerated, leading to a faster decrease of $T_e$ and a faster increase of $T_i$. It is worthwhile to mention here, that temperature change of the ions is more drastic, compared to the electrons. While the electron temperature stays at the same order of magnitude the ion temperature increases by a factor of 50.

We performed conductivity calculations along the path of the evolution of $T_e$ and $T_i$  for different laser energy densities. For this purpose, we first carried out a DFT-MD simulation with 108 atoms at different sets of fixed $T_e$ and $T_i$ which correspond to a point in time of the relaxation process. For a number of 10 snapshots from these simulations  we performed conductivity calculations applying the Kubo-Greenwood formula.  

\begin{figure}[htb]
 \includegraphics[width=0.6\columnwidth]{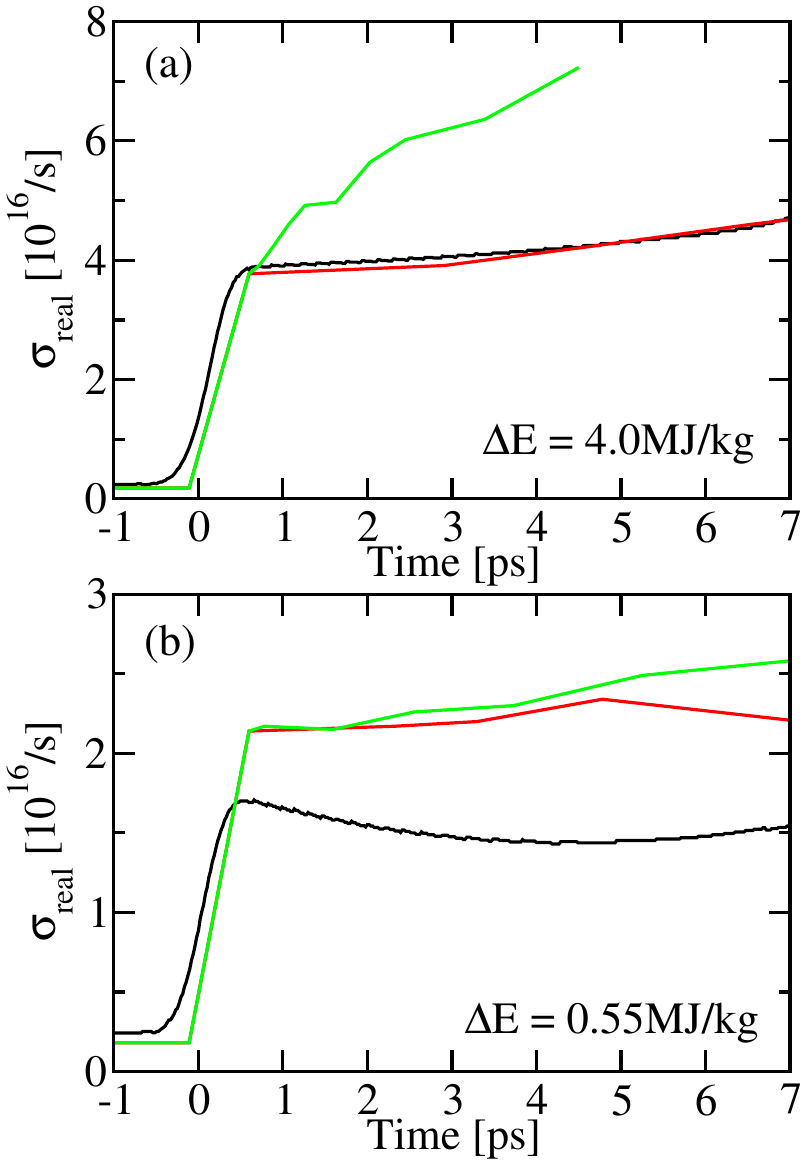}
 \caption{Evolution of the AC conductivity of gold for a laser energy density of 4.0~MJ/kg (a)  and 0.55~MJ/kg (b). Values from experiment are shown in black, the constant $G_{ei}$ model in red and the $T_e$-dependent $G_{ei}$ model in green.}
 \label{fig:sigma-evol}
\end{figure}

In Fig.~\ref{fig:sigma-evol} we compare the model findings with experimental data for laser energy densities of 0.55~MJ/kg and 4.0~MJ/kg. Two curves for each case were calculated, one using $G_{ei}$ calculated as described above and another one using a constant coupling factor as found in previous experiment at low laser energy density.~\cite{Hohlfeld2000}

The AC conductivity evolution at a  photon energy of 1.55~eV is dominated by two competing processes: a decrease while the electron temperature is declining and an increase that is caused by a growing ion temperature. The DC conductivity is decreasing with higher ion temperature, which is usually explained by an increased perturbation caused by the higher amplitude of the vibrations of the ions around their equilibrium positions. This mechanism is not observed for the AC conductivity at a photon energy of 1.55~eV. Here the minimum in $\sigma_r$, which can be observed in Fig.~\ref{fig:sigma300K}(a) vanishes and consequently the AC conductivity is increasing. 

The electron temperature decreases with time, as the electrons lose their kinetic energy to the ions.
Consequently also a decreasing conductivity could be expected, which is true at an energy density of  0.55~MJ/kg but is not observed for 4.0~MJ/kg. The higher energy density causes the ions to be heated strong enough, so that the resulting change in conductivity can compensate the effect of decreasing electron temperature. As a result the conductivity increases. The relative dominance  of either of these two processes is governed by $G_{ei}$.

The evolution of the AC conductivity at 4.0~MJ/kg in Fig.~\ref{fig:sigma-evol}(a) indicates that the model curve agrees better with the measurements if the constant value of $G_{ei}=2.2\times10^{16}$~W/m$^3$K is used instead of the temperature-dependent electron-phonon coupling constant. A consequence from that observation might be that the electron-phonon coupling factor is smaller than predicted by the method of electron-phonon coupling explained above. 

In a recent approach,~\cite{Mueller2013} the electron-phonon coupling was found to have a similar temperature dependence as presented here.  This suggests, that the uncertainties within the method itself seem to be well controlled. All methods, that are implementing electron-phonon coupling so far, do not include the ion temperature into the calculation of $\alpha^2F(\Omega)$ and thus the constant $\lambda\langle\omega^2\rangle$ is completely independent from temperature. To our knowledge, there have been no successful attempts to include the lattice temperature in calculations of the electron-phonon coupling. Electron-ion coupling was also described for plasmas as a scattering of electrons at plasmons, which became known as coupled mode calculations. For aluminum this method results in substantially lower coupling constants than resulting from the electron-phonon scattering approach.~\cite{Dharma-wardana1998,Vorberger2012}

At the energy density of 0.55~MJ/kg the reached electron temperature does not exceed $12,000$~K. In that temperature range the $G_{ei}$ obtained by \textsc{Abinit} is not substantially higher than the constant one, see Fig.~\ref{fig:gei}. As a result, the temperature evolution and consequently the AC conductivity evolution is very similar for the both cases and is in good agreement with the measured data, see Fig.~\ref{fig:sigma-evol}(b). However, there is a small difference in the absolute value, which is already be seen in the initial conductivity at 0.55~MJ/kg plotted in Fig.~\ref{fig:initial-cond}(a).

\section{Conclusions}
\label{sec:conclusion}

We describe the  thermophysical properties of two-temperature warm dense gold using an \textit{ab initio} model. DFT simulations allowed the calculation of thermodynamic parameters and the optical conductivity for two-temperature states within the same theoretical framework. 

Our results for the electron heat capacity have been corroborated by comparison with experimental data for the initial conductivity obtained in thin gold foils.~\cite{Chen2013} We could demonstrate that the real part of the AC conductivity shows reasonable agreement with the measured data during the temperature relaxation process. Convincing agreement with experimental result for the imaginary part is not observed. We identified the insufficient precision in our calculation of intraband contributions to the AC conductivity of metals, which leads to an underestimation of the static limit of the real part as the main source of error for this behavior. This causes errors within the application of the Kramers-Kronig relation, while the real part at the experimental laser frequency is not effected and is in agreement with the measurements.

We showed, that the model can be used in combination with experimental data to investigate the material properties, namely the electron heat capacity and the electron-ion coupling factor. We found the present \textit{ab initio} calculations for the electron heat capacity of warm dense gold to be sufficient. A comparison of experimental results clearly shows, that the electron-ion energy transfer rate at high laser energy density is overestimated by current electron-phonon coupling models.~\cite{Chen2013}

The present work can be easily applied to other metals and can be used to investigate their material properties at warm dense matter conditions. To improve the conductivity calculations, we propose on the one hand the increase of the atom number, which will improve the calculation of intraband contributions and therefore lead to a more reasonable result also for the imaginary part. On the other hand, the use of an improved exchange-correlation functional might fix the shift against the measured conductivity at 300~K. 

A major remaining issue is the behavior of the electron-ion coupling for the conditions considered here. Experiments suggest that the coupling factor is likely to have a near constant behavior and lower than predicted by our model. The inclusion of ion temperature into the calculation of the electron-phonon coupling might be the answer to that problem. Another possible solution might be the modelling of the energy transfer as electron-plasmon scattering, which is implemented in the coupled-mode approach.~\cite{Dharma-wardana1998, Vorberger2012}

\begin{acknowledgments}
We would like to thank Patrick Audebert, Patrick Renaudin and Arnaud Sollier for inspiring discussions. We received financial support from the French Agence Nationale de la Recherche, project OEDYP, Grant No. ANR-09-BLAN-0206. Part of the numerical calculations have been performed using resources from GENCI under Grant No. 6454. This work is supported by the Natural Sciences \& Engineering Research Council of Canada.
\end{acknowledgments}


\bibliography{gold}

\begin{thebibliography}{64}
\expandafter\ifx\csname natexlab\endcsname\relax\def\natexlab#1{#1}\fi
\expandafter\ifx\csname bibnamefont\endcsname\relax
  \def\bibnamefont#1{#1}\fi
\expandafter\ifx\csname bibfnamefont\endcsname\relax
  \def\bibfnamefont#1{#1}\fi
\expandafter\ifx\csname citenamefont\endcsname\relax
  \def\citenamefont#1{#1}\fi
\expandafter\ifx\csname url\endcsname\relax
  \def\url#1{\texttt{#1}}\fi
\expandafter\ifx\csname urlprefix\endcsname\relax\def\urlprefix{URL }\fi
\providecommand{\bibinfo}[2]{#2}
\providecommand{\eprint}[2][]{\url{#2}}

\bibitem[{\citenamefont{Chen et~al.}(2013)\citenamefont{Chen, Holst, Kirkwood,
  Sametoglu, Reid, Tsui, Recoules, and Ng}}]{Chen2013}
\bibinfo{author}{\bibfnamefont{Z.}~\bibnamefont{Chen}},
  \bibinfo{author}{\bibfnamefont{B.}~\bibnamefont{Holst}},
  \bibinfo{author}{\bibfnamefont{S.~E.} \bibnamefont{Kirkwood}},
  \bibinfo{author}{\bibfnamefont{V.}~\bibnamefont{Sametoglu}},
  \bibinfo{author}{\bibfnamefont{M.}~\bibnamefont{Reid}},
  \bibinfo{author}{\bibfnamefont{Y.~Y.} \bibnamefont{Tsui}},
  \bibinfo{author}{\bibfnamefont{V.}~\bibnamefont{Recoules}}, \bibnamefont{and}
  \bibinfo{author}{\bibfnamefont{A.}~\bibnamefont{Ng}}, \bibinfo{journal}{Phys.
  Rev. Lett.} \textbf{\bibinfo{volume}{110}}, \bibinfo{pages}{135001}
  (\bibinfo{year}{2013}).

\bibitem[{\citenamefont{Sokolowski-Tinten
  et~al.}(2003)\citenamefont{Sokolowski-Tinten, Blome, Blums, Cavalleri,
  Dietrich, Tarasevitch, Uschmann, Forster, Kammler, Horn-von Hoegen
  et~al.}}]{Sokolowski-Tinten2003}
\bibinfo{author}{\bibfnamefont{K.}~\bibnamefont{Sokolowski-Tinten}},
  \bibinfo{author}{\bibfnamefont{C.}~\bibnamefont{Blome}},
  \bibinfo{author}{\bibfnamefont{J.}~\bibnamefont{Blums}},
  \bibinfo{author}{\bibfnamefont{A.}~\bibnamefont{Cavalleri}},
  \bibinfo{author}{\bibfnamefont{C.}~\bibnamefont{Dietrich}},
  \bibinfo{author}{\bibfnamefont{A.}~\bibnamefont{Tarasevitch}},
  \bibinfo{author}{\bibfnamefont{I.}~\bibnamefont{Uschmann}},
  \bibinfo{author}{\bibfnamefont{E.}~\bibnamefont{Forster}},
  \bibinfo{author}{\bibfnamefont{M.}~\bibnamefont{Kammler}},
  \bibinfo{author}{\bibfnamefont{M.}~\bibnamefont{Horn-von Hoegen}},
  \bibnamefont{et~al.}, \bibinfo{journal}{Nature}
  \textbf{\bibinfo{volume}{422}}, \bibinfo{pages}{287} (\bibinfo{year}{2003}).

\bibitem[{\citenamefont{Uteza et~al.}(2004)\citenamefont{Uteza, Gamaly, Rode,
  Samoc, and Luther-Davies}}]{Uteza2004}
\bibinfo{author}{\bibfnamefont{O.~P.} \bibnamefont{Uteza}},
  \bibinfo{author}{\bibfnamefont{E.~G.} \bibnamefont{Gamaly}},
  \bibinfo{author}{\bibfnamefont{A.~V.} \bibnamefont{Rode}},
  \bibinfo{author}{\bibfnamefont{M.}~\bibnamefont{Samoc}}, \bibnamefont{and}
  \bibinfo{author}{\bibfnamefont{B.}~\bibnamefont{Luther-Davies}},
  \bibinfo{journal}{Phys. Rev. B} \textbf{\bibinfo{volume}{70}},
  \bibinfo{pages}{054108} (\bibinfo{year}{2004}).

\bibitem[{\citenamefont{Widmann et~al.}(2004)\citenamefont{Widmann, Ao, Foord,
  Price, Ellis, Springer, and Ng}}]{Widmann2004}
\bibinfo{author}{\bibfnamefont{K.}~\bibnamefont{Widmann}},
  \bibinfo{author}{\bibfnamefont{T.}~\bibnamefont{Ao}},
  \bibinfo{author}{\bibfnamefont{M.~E.} \bibnamefont{Foord}},
  \bibinfo{author}{\bibfnamefont{D.~F.} \bibnamefont{Price}},
  \bibinfo{author}{\bibfnamefont{A.~D.} \bibnamefont{Ellis}},
  \bibinfo{author}{\bibfnamefont{P.~T.} \bibnamefont{Springer}},
  \bibnamefont{and} \bibinfo{author}{\bibfnamefont{A.}~\bibnamefont{Ng}},
  \bibinfo{journal}{Phys. Rev. Lett.} \textbf{\bibinfo{volume}{92}},
  \bibinfo{pages}{125002} (\bibinfo{year}{2004}).

\bibitem[{\citenamefont{Ping et~al.}(2006)\citenamefont{Ping, Hanson, Koslow,
  Ogitsu, Prendergast, Schwegler, Collins, and Ng}}]{Ping2006}
\bibinfo{author}{\bibfnamefont{Y.}~\bibnamefont{Ping}},
  \bibinfo{author}{\bibfnamefont{D.}~\bibnamefont{Hanson}},
  \bibinfo{author}{\bibfnamefont{I.}~\bibnamefont{Koslow}},
  \bibinfo{author}{\bibfnamefont{T.}~\bibnamefont{Ogitsu}},
  \bibinfo{author}{\bibfnamefont{D.}~\bibnamefont{Prendergast}},
  \bibinfo{author}{\bibfnamefont{E.}~\bibnamefont{Schwegler}},
  \bibinfo{author}{\bibfnamefont{G.}~\bibnamefont{Collins}}, \bibnamefont{and}
  \bibinfo{author}{\bibfnamefont{A.}~\bibnamefont{Ng}}, \bibinfo{journal}{Phys.
  Rev. Lett.} \textbf{\bibinfo{volume}{96}}, \bibinfo{pages}{255003}
  (\bibinfo{year}{2006}).

\bibitem[{\citenamefont{Ao et~al.}(2006)\citenamefont{Ao, Ping, Widmann, Price,
  Lee, Tam, Springer, and Ng}}]{Ao2006}
\bibinfo{author}{\bibfnamefont{T.}~\bibnamefont{Ao}},
  \bibinfo{author}{\bibfnamefont{Y.}~\bibnamefont{Ping}},
  \bibinfo{author}{\bibfnamefont{K.}~\bibnamefont{Widmann}},
  \bibinfo{author}{\bibfnamefont{D.~F.} \bibnamefont{Price}},
  \bibinfo{author}{\bibfnamefont{E.}~\bibnamefont{Lee}},
  \bibinfo{author}{\bibfnamefont{H.}~\bibnamefont{Tam}},
  \bibinfo{author}{\bibfnamefont{P.~T.} \bibnamefont{Springer}},
  \bibnamefont{and} \bibinfo{author}{\bibfnamefont{A.}~\bibnamefont{Ng}},
  \bibinfo{journal}{Phys. Rev. Lett.} \textbf{\bibinfo{volume}{96}},
  \bibinfo{pages}{055001} (\bibinfo{year}{2006}).

\bibitem[{\citenamefont{Kandyla et~al.}(2007)\citenamefont{Kandyla, Shih, and
  Mazur}}]{Kandyla2007}
\bibinfo{author}{\bibfnamefont{M.}~\bibnamefont{Kandyla}},
  \bibinfo{author}{\bibfnamefont{T.}~\bibnamefont{Shih}}, \bibnamefont{and}
  \bibinfo{author}{\bibfnamefont{E.}~\bibnamefont{Mazur}},
  \bibinfo{journal}{Phys. Rev. B} \textbf{\bibinfo{volume}{75}},
  \bibinfo{pages}{214107} (\bibinfo{year}{2007}).

\bibitem[{\citenamefont{Ping et~al.}(2008)\citenamefont{Ping, Hanson, Koslow,
  Ogitsu, Prendergast, Schwegler, Collins, and Ng}}]{Ping2008}
\bibinfo{author}{\bibfnamefont{Y.}~\bibnamefont{Ping}},
  \bibinfo{author}{\bibfnamefont{D.}~\bibnamefont{Hanson}},
  \bibinfo{author}{\bibfnamefont{I.}~\bibnamefont{Koslow}},
  \bibinfo{author}{\bibfnamefont{T.}~\bibnamefont{Ogitsu}},
  \bibinfo{author}{\bibfnamefont{D.}~\bibnamefont{Prendergast}},
  \bibinfo{author}{\bibfnamefont{E.}~\bibnamefont{Schwegler}},
  \bibinfo{author}{\bibfnamefont{G.}~\bibnamefont{Collins}}, \bibnamefont{and}
  \bibinfo{author}{\bibfnamefont{A.}~\bibnamefont{Ng}}, \bibinfo{journal}{Phys.
  Plasmas} \textbf{\bibinfo{volume}{15}}, \bibinfo{eid}{056303}
  (\bibinfo{year}{2008}).

\bibitem[{\citenamefont{Ernstorfer et~al.}(2009)\citenamefont{Ernstorfer, Harb,
  Hebeisen, Sciaini, Dartigalongue, and Miller}}]{Ernstorfer2009}
\bibinfo{author}{\bibfnamefont{R.}~\bibnamefont{Ernstorfer}},
  \bibinfo{author}{\bibfnamefont{M.}~\bibnamefont{Harb}},
  \bibinfo{author}{\bibfnamefont{C.~T.} \bibnamefont{Hebeisen}},
  \bibinfo{author}{\bibfnamefont{G.}~\bibnamefont{Sciaini}},
  \bibinfo{author}{\bibfnamefont{T.}~\bibnamefont{Dartigalongue}},
  \bibnamefont{and} \bibinfo{author}{\bibfnamefont{R.~J.~D.}
  \bibnamefont{Miller}}, \bibinfo{journal}{Science}
  \textbf{\bibinfo{volume}{323}}, \bibinfo{pages}{1033} (\bibinfo{year}{2009}).

\bibitem[{\citenamefont{Ping et~al.}(2010)\citenamefont{Ping, Correa, Ogitsu,
  Draeger, Schwegler, Ao, Widmann, Price, Lee, Tam et~al.}}]{Ping2010}
\bibinfo{author}{\bibfnamefont{Y.}~\bibnamefont{Ping}},
  \bibinfo{author}{\bibfnamefont{A.}~\bibnamefont{Correa}},
  \bibinfo{author}{\bibfnamefont{T.}~\bibnamefont{Ogitsu}},
  \bibinfo{author}{\bibfnamefont{E.}~\bibnamefont{Draeger}},
  \bibinfo{author}{\bibfnamefont{E.}~\bibnamefont{Schwegler}},
  \bibinfo{author}{\bibfnamefont{T.}~\bibnamefont{Ao}},
  \bibinfo{author}{\bibfnamefont{K.}~\bibnamefont{Widmann}},
  \bibinfo{author}{\bibfnamefont{D.}~\bibnamefont{Price}},
  \bibinfo{author}{\bibfnamefont{E.}~\bibnamefont{Lee}},
  \bibinfo{author}{\bibfnamefont{H.}~\bibnamefont{Tam}}, \bibnamefont{et~al.},
  \bibinfo{journal}{High Energy Density Phys.} \textbf{\bibinfo{volume}{6}},
  \bibinfo{pages}{246 } (\bibinfo{year}{2010}).

\bibitem[{\citenamefont{Cho et~al.}(2011)\citenamefont{Cho, Engelhorn, Correa,
  Ogitsu, Weber, Lee, Feng, Ni, Ping, Nelson et~al.}}]{Cho2011}
\bibinfo{author}{\bibfnamefont{B.~I.} \bibnamefont{Cho}},
  \bibinfo{author}{\bibfnamefont{K.}~\bibnamefont{Engelhorn}},
  \bibinfo{author}{\bibfnamefont{A.~A.} \bibnamefont{Correa}},
  \bibinfo{author}{\bibfnamefont{T.}~\bibnamefont{Ogitsu}},
  \bibinfo{author}{\bibfnamefont{C.~P.} \bibnamefont{Weber}},
  \bibinfo{author}{\bibfnamefont{H.~J.} \bibnamefont{Lee}},
  \bibinfo{author}{\bibfnamefont{J.}~\bibnamefont{Feng}},
  \bibinfo{author}{\bibfnamefont{P.~A.} \bibnamefont{Ni}},
  \bibinfo{author}{\bibfnamefont{Y.}~\bibnamefont{Ping}},
  \bibinfo{author}{\bibfnamefont{A.~J.} \bibnamefont{Nelson}},
  \bibnamefont{et~al.}, \bibinfo{journal}{Phys. Rev. Lett.}
  \textbf{\bibinfo{volume}{106}}, \bibinfo{pages}{167601}
  (\bibinfo{year}{2011}).

\bibitem[{\citenamefont{Fann et~al.}(1992{\natexlab{a}})\citenamefont{Fann,
  Storz, Tom, and Bokor}}]{Fann1992}
\bibinfo{author}{\bibfnamefont{W.~S.} \bibnamefont{Fann}},
  \bibinfo{author}{\bibfnamefont{R.}~\bibnamefont{Storz}},
  \bibinfo{author}{\bibfnamefont{H.~W.~K.} \bibnamefont{Tom}},
  \bibnamefont{and} \bibinfo{author}{\bibfnamefont{J.}~\bibnamefont{Bokor}},
  \bibinfo{journal}{Phys. Rev. B} \textbf{\bibinfo{volume}{46}},
  \bibinfo{pages}{13592} (\bibinfo{year}{1992}{\natexlab{a}}).

\bibitem[{\citenamefont{Fann et~al.}(1992{\natexlab{b}})\citenamefont{Fann,
  Storz, Tom, and Bokor}}]{Fann1992a}
\bibinfo{author}{\bibfnamefont{W.~S.} \bibnamefont{Fann}},
  \bibinfo{author}{\bibfnamefont{R.}~\bibnamefont{Storz}},
  \bibinfo{author}{\bibfnamefont{H.~W.~K.} \bibnamefont{Tom}},
  \bibnamefont{and} \bibinfo{author}{\bibfnamefont{J.}~\bibnamefont{Bokor}},
  \bibinfo{journal}{Phys. Rev. Lett.} \textbf{\bibinfo{volume}{68}},
  \bibinfo{pages}{2834} (\bibinfo{year}{1992}{\natexlab{b}}).

\bibitem[{\citenamefont{Hohlfeld et~al.}(2000)\citenamefont{Hohlfeld,
  Wellershoff, G\"udde, Conrad, J\"ahnke, and Matthias}}]{Hohlfeld2000}
\bibinfo{author}{\bibfnamefont{J.}~\bibnamefont{Hohlfeld}},
  \bibinfo{author}{\bibfnamefont{S.-S.} \bibnamefont{Wellershoff}},
  \bibinfo{author}{\bibfnamefont{J.}~\bibnamefont{G\"udde}},
  \bibinfo{author}{\bibfnamefont{U.}~\bibnamefont{Conrad}},
  \bibinfo{author}{\bibfnamefont{V.}~\bibnamefont{J\"ahnke}}, \bibnamefont{and}
  \bibinfo{author}{\bibfnamefont{E.}~\bibnamefont{Matthias}},
  \bibinfo{journal}{Chem. Phys.} \textbf{\bibinfo{volume}{251}},
  \bibinfo{pages}{237 } (\bibinfo{year}{2000}).

\bibitem[{\citenamefont{Bonn et~al.}(2000)\citenamefont{Bonn, Denzler, Funk,
  Wolf, Wellershoff, and Hohlfeld}}]{Bonn2000}
\bibinfo{author}{\bibfnamefont{M.}~\bibnamefont{Bonn}},
  \bibinfo{author}{\bibfnamefont{D.~N.} \bibnamefont{Denzler}},
  \bibinfo{author}{\bibfnamefont{S.}~\bibnamefont{Funk}},
  \bibinfo{author}{\bibfnamefont{M.}~\bibnamefont{Wolf}},
  \bibinfo{author}{\bibfnamefont{S.-S.} \bibnamefont{Wellershoff}},
  \bibnamefont{and} \bibinfo{author}{\bibfnamefont{J.}~\bibnamefont{Hohlfeld}},
  \bibinfo{journal}{Phys. Rev. B} \textbf{\bibinfo{volume}{61}},
  \bibinfo{pages}{1101} (\bibinfo{year}{2000}).

\bibitem[{\citenamefont{Hohlfeld et~al.}(1997)\citenamefont{Hohlfeld, M\"uller,
  Wellershoff, and Matthias}}]{Hohlfeld1997}
\bibinfo{author}{\bibfnamefont{J.}~\bibnamefont{Hohlfeld}},
  \bibinfo{author}{\bibfnamefont{J.}~\bibnamefont{M\"uller}},
  \bibinfo{author}{\bibfnamefont{S.-S.} \bibnamefont{Wellershoff}},
  \bibnamefont{and} \bibinfo{author}{\bibfnamefont{E.}~\bibnamefont{Matthias}},
  \bibinfo{journal}{Appl. Phys. B: Lasers Opt.} \textbf{\bibinfo{volume}{64}},
  \bibinfo{pages}{387} (\bibinfo{year}{1997}).

\bibitem[{\citenamefont{Su\'arez et~al.}(1995)\citenamefont{Su\'arez, Bron, and
  Juhasz}}]{Suarez1995}
\bibinfo{author}{\bibfnamefont{C.}~\bibnamefont{Su\'arez}},
  \bibinfo{author}{\bibfnamefont{W.~E.} \bibnamefont{Bron}}, \bibnamefont{and}
  \bibinfo{author}{\bibfnamefont{T.}~\bibnamefont{Juhasz}},
  \bibinfo{journal}{Phys. Rev. Lett.} \textbf{\bibinfo{volume}{75}},
  \bibinfo{pages}{4536} (\bibinfo{year}{1995}).

\bibitem[{\citenamefont{Juhasz et~al.}(1993)\citenamefont{Juhasz, Elsayed-Ali,
  Smith, Su\'arez, and Bron}}]{Juhasz1993}
\bibinfo{author}{\bibfnamefont{T.}~\bibnamefont{Juhasz}},
  \bibinfo{author}{\bibfnamefont{H.~E.} \bibnamefont{Elsayed-Ali}},
  \bibinfo{author}{\bibfnamefont{G.~O.} \bibnamefont{Smith}},
  \bibinfo{author}{\bibfnamefont{C.}~\bibnamefont{Su\'arez}}, \bibnamefont{and}
  \bibinfo{author}{\bibfnamefont{W.~E.} \bibnamefont{Bron}},
  \bibinfo{journal}{Phys. Rev. B} \textbf{\bibinfo{volume}{48}},
  \bibinfo{pages}{15488} (\bibinfo{year}{1993}).

\bibitem[{\citenamefont{Anisimov et~al.}(1974)\citenamefont{Anisimov, L., and
  Perelman}}]{Anisimov1974a}
\bibinfo{author}{\bibfnamefont{S.~I.} \bibnamefont{Anisimov}},
  \bibinfo{author}{\bibfnamefont{K.~B.} \bibnamefont{L.}}, \bibnamefont{and}
  \bibinfo{author}{\bibfnamefont{T.~L.} \bibnamefont{Perelman}},
  \bibinfo{journal}{Sov. Phys. JETP} \textbf{\bibinfo{volume}{39}},
  \bibinfo{pages}{375} (\bibinfo{year}{1974}).

\bibitem[{\citenamefont{Mazevet et~al.}(2005)\citenamefont{Mazevet, Cl\'erouin,
  Recoules, Anglade, and Z\'erah}}]{Mazevet2005}
\bibinfo{author}{\bibfnamefont{S.}~\bibnamefont{Mazevet}},
  \bibinfo{author}{\bibfnamefont{J.}~\bibnamefont{Cl\'erouin}},
  \bibinfo{author}{\bibfnamefont{V.}~\bibnamefont{Recoules}},
  \bibinfo{author}{\bibfnamefont{P.~M.} \bibnamefont{Anglade}},
  \bibnamefont{and} \bibinfo{author}{\bibfnamefont{G.}~\bibnamefont{Z\'erah}},
  \bibinfo{journal}{Phys. Rev. Lett.} \textbf{\bibinfo{volume}{95}},
  \bibinfo{pages}{085002} (\bibinfo{year}{2005}).

\bibitem[{\citenamefont{Allen}(1987)}]{Allen1987}
\bibinfo{author}{\bibfnamefont{P.~B.} \bibnamefont{Allen}},
  \bibinfo{journal}{Phys. Rev. Lett.} \textbf{\bibinfo{volume}{59}},
  \bibinfo{pages}{1460} (\bibinfo{year}{1987}).

\bibitem[{\citenamefont{Wang et~al.}(1994)\citenamefont{Wang, Riffe, Lee, and
  Downer}}]{Wang1994}
\bibinfo{author}{\bibfnamefont{X.~Y.} \bibnamefont{Wang}},
  \bibinfo{author}{\bibfnamefont{D.~M.} \bibnamefont{Riffe}},
  \bibinfo{author}{\bibfnamefont{Y.-S.} \bibnamefont{Lee}}, \bibnamefont{and}
  \bibinfo{author}{\bibfnamefont{M.~C.} \bibnamefont{Downer}},
  \bibinfo{journal}{Phys. Rev. B} \textbf{\bibinfo{volume}{50}},
  \bibinfo{pages}{8016} (\bibinfo{year}{1994}).

\bibitem[{\citenamefont{Lin et~al.}(2008)\citenamefont{Lin, Zhigilei, and
  Celli}}]{Lin2008}
\bibinfo{author}{\bibfnamefont{Z.}~\bibnamefont{Lin}},
  \bibinfo{author}{\bibfnamefont{L.~V.} \bibnamefont{Zhigilei}},
  \bibnamefont{and} \bibinfo{author}{\bibfnamefont{V.}~\bibnamefont{Celli}},
  \bibinfo{journal}{Phys. Rev. B} \textbf{\bibinfo{volume}{77}},
  \bibinfo{pages}{075133} (\bibinfo{year}{2008}).

\bibitem[{\citenamefont{Dharma-wardana and Perrot}(1998)}]{Dharma-wardana1998}
\bibinfo{author}{\bibfnamefont{M.~W.~C.} \bibnamefont{Dharma-wardana}}
  \bibnamefont{and} \bibinfo{author}{\bibfnamefont{F.}~\bibnamefont{Perrot}},
  \bibinfo{journal}{Phys. Rev. E} \textbf{\bibinfo{volume}{58}},
  \bibinfo{pages}{3705} (\bibinfo{year}{1998}).

\bibitem[{\citenamefont{Vorberger and Gericke}(2012)}]{Vorberger2012}
\bibinfo{author}{\bibfnamefont{J.}~\bibnamefont{Vorberger}} \bibnamefont{and}
  \bibinfo{author}{\bibfnamefont{D.~O.} \bibnamefont{Gericke}},
  \bibinfo{journal}{AIP Conf. Proc.} \textbf{\bibinfo{volume}{1464}},
  \bibinfo{pages}{572} (\bibinfo{year}{2012}).

\bibitem[{\citenamefont{Spitzer and H\"arm}(1953)}]{Spitzer1953}
\bibinfo{author}{\bibfnamefont{L.}~\bibnamefont{Spitzer}} \bibnamefont{and}
  \bibinfo{author}{\bibfnamefont{R.}~\bibnamefont{H\"arm}},
  \bibinfo{journal}{Phys. Rev.} \textbf{\bibinfo{volume}{89}},
  \bibinfo{pages}{977} (\bibinfo{year}{1953}).

\bibitem[{\citenamefont{Lee and More}(1984)}]{Lee1984}
\bibinfo{author}{\bibfnamefont{Y.~T.} \bibnamefont{Lee}} \bibnamefont{and}
  \bibinfo{author}{\bibfnamefont{R.~M.} \bibnamefont{More}},
  \bibinfo{journal}{Phys. Fluids} \textbf{\bibinfo{volume}{27}},
  \bibinfo{pages}{1273} (\bibinfo{year}{1984}).

\bibitem[{\citenamefont{Ziman}(1961)}]{Ziman1961}
\bibinfo{author}{\bibfnamefont{J.~M.} \bibnamefont{Ziman}},
  \bibinfo{journal}{Philos. Mag.} \textbf{\bibinfo{volume}{6}},
  \bibinfo{pages}{1013} (\bibinfo{year}{1961}).

\bibitem[{\citenamefont{Rinker}(1985)}]{Rinker1985}
\bibinfo{author}{\bibfnamefont{G.~A.} \bibnamefont{Rinker}},
  \bibinfo{journal}{Phys. Rev. B} \textbf{\bibinfo{volume}{31}},
  \bibinfo{pages}{4207} (\bibinfo{year}{1985}).

\bibitem[{\citenamefont{Rinker}(1988)}]{Rinker1988}
\bibinfo{author}{\bibfnamefont{G.~A.} \bibnamefont{Rinker}},
  \bibinfo{journal}{Phys. Rev. A} \textbf{\bibinfo{volume}{37}},
  \bibinfo{pages}{1284} (\bibinfo{year}{1988}).

\bibitem[{\citenamefont{Perrot and Dharma-wardana}(1987)}]{Perrot1987}
\bibinfo{author}{\bibfnamefont{F.}~\bibnamefont{Perrot}} \bibnamefont{and}
  \bibinfo{author}{\bibfnamefont{M.~W.~C.} \bibnamefont{Dharma-wardana}},
  \bibinfo{journal}{Phys. Rev. A} \textbf{\bibinfo{volume}{36}},
  \bibinfo{pages}{238} (\bibinfo{year}{1987}).

\bibitem[{\citenamefont{Perrot and Dharma-wardana}(1995)}]{Perrot1995}
\bibinfo{author}{\bibfnamefont{F.}~\bibnamefont{Perrot}} \bibnamefont{and}
  \bibinfo{author}{\bibfnamefont{M.~W.~C.} \bibnamefont{Dharma-wardana}},
  \bibinfo{journal}{Phys. Rev. E} \textbf{\bibinfo{volume}{52}},
  \bibinfo{pages}{5352} (\bibinfo{year}{1995}).

\bibitem[{\citenamefont{Dharma-wardana}(2006)}]{Dharma-wardana2006}
\bibinfo{author}{\bibfnamefont{M.~W.~C.} \bibnamefont{Dharma-wardana}},
  \bibinfo{journal}{Phys. Rev. E} \textbf{\bibinfo{volume}{73}},
  \bibinfo{pages}{036401} (\bibinfo{year}{2006}).

\bibitem[{\citenamefont{DeSilva and Katsouros}(1998)}]{DeSilva1998}
\bibinfo{author}{\bibfnamefont{A.~W.} \bibnamefont{DeSilva}} \bibnamefont{and}
  \bibinfo{author}{\bibfnamefont{J.~D.} \bibnamefont{Katsouros}},
  \bibinfo{journal}{Phys. Rev. E} \textbf{\bibinfo{volume}{57}},
  \bibinfo{pages}{5945} (\bibinfo{year}{1998}).

\bibitem[{\citenamefont{Desjarlais et~al.}(2002)\citenamefont{Desjarlais,
  Kress, and Collins}}]{Desjarlais2002}
\bibinfo{author}{\bibfnamefont{M.~P.} \bibnamefont{Desjarlais}},
  \bibinfo{author}{\bibfnamefont{J.~D.} \bibnamefont{Kress}}, \bibnamefont{and}
  \bibinfo{author}{\bibfnamefont{L.~A.} \bibnamefont{Collins}},
  \bibinfo{journal}{Phys. Rev. E} \textbf{\bibinfo{volume}{66}},
  \bibinfo{pages}{025401} (\bibinfo{year}{2002}).

\bibitem[{\citenamefont{Vast et~al.}(1995)\citenamefont{Vast, Bernard, and
  Zerah}}]{Vast1995}
\bibinfo{author}{\bibfnamefont{N.}~\bibnamefont{Vast}},
  \bibinfo{author}{\bibfnamefont{S.}~\bibnamefont{Bernard}}, \bibnamefont{and}
  \bibinfo{author}{\bibfnamefont{G.}~\bibnamefont{Zerah}},
  \bibinfo{journal}{Phys. Rev. B} \textbf{\bibinfo{volume}{52}},
  \bibinfo{pages}{4123} (\bibinfo{year}{1995}).

\bibitem[{\citenamefont{Holender et~al.}(1995)\citenamefont{Holender, Gillan,
  Payne, and Simpson}}]{Holender1995}
\bibinfo{author}{\bibfnamefont{J.~M.} \bibnamefont{Holender}},
  \bibinfo{author}{\bibfnamefont{M.~J.} \bibnamefont{Gillan}},
  \bibinfo{author}{\bibfnamefont{M.~C.} \bibnamefont{Payne}}, \bibnamefont{and}
  \bibinfo{author}{\bibfnamefont{A.~D.} \bibnamefont{Simpson}},
  \bibinfo{journal}{Phys. Rev. B} \textbf{\bibinfo{volume}{52}},
  \bibinfo{pages}{967} (\bibinfo{year}{1995}).

\bibitem[{\citenamefont{Silvestrelli}(1999)}]{Silvestrelli1999}
\bibinfo{author}{\bibfnamefont{P.~L.} \bibnamefont{Silvestrelli}},
  \bibinfo{journal}{Phys. Rev. B} \textbf{\bibinfo{volume}{60}},
  \bibinfo{pages}{16382} (\bibinfo{year}{1999}).

\bibitem[{\citenamefont{White et~al.}(2012)\citenamefont{White, Vorberger,
  Brown, Crowley, Davis, Glenzer, Harris, Hochhaus, Le~Pape, Ma
  et~al.}}]{White2012}
\bibinfo{author}{\bibfnamefont{T.~G.} \bibnamefont{White}},
  \bibinfo{author}{\bibfnamefont{J.}~\bibnamefont{Vorberger}},
  \bibinfo{author}{\bibfnamefont{C.~R.~D.} \bibnamefont{Brown}},
  \bibinfo{author}{\bibfnamefont{B.~J.~B.} \bibnamefont{Crowley}},
  \bibinfo{author}{\bibfnamefont{P.}~\bibnamefont{Davis}},
  \bibinfo{author}{\bibfnamefont{S.~H.} \bibnamefont{Glenzer}},
  \bibinfo{author}{\bibfnamefont{J.~W.~O.} \bibnamefont{Harris}},
  \bibinfo{author}{\bibfnamefont{D.~C.} \bibnamefont{Hochhaus}},
  \bibinfo{author}{\bibfnamefont{S.}~\bibnamefont{Le~Pape}},
  \bibinfo{author}{\bibfnamefont{T.}~\bibnamefont{Ma}}, \bibnamefont{et~al.},
  \bibinfo{journal}{Scientific Reports} \textbf{\bibinfo{volume}{2}},
  \bibinfo{pages}{889} (\bibinfo{year}{2012}).

\bibitem[{\citenamefont{Gonze et~al.}(2009)\citenamefont{Gonze, Amadon,
  Anglade, Beuken, Bottin, Boulanger, Bruneval, Caliste, Caracas, C\^ot\'e
  et~al.}}]{Gonze2009}
\bibinfo{author}{\bibfnamefont{X.}~\bibnamefont{Gonze}},
  \bibinfo{author}{\bibfnamefont{B.}~\bibnamefont{Amadon}},
  \bibinfo{author}{\bibfnamefont{P.-M.} \bibnamefont{Anglade}},
  \bibinfo{author}{\bibfnamefont{J.-M.} \bibnamefont{Beuken}},
  \bibinfo{author}{\bibfnamefont{F.}~\bibnamefont{Bottin}},
  \bibinfo{author}{\bibfnamefont{P.}~\bibnamefont{Boulanger}},
  \bibinfo{author}{\bibfnamefont{F.}~\bibnamefont{Bruneval}},
  \bibinfo{author}{\bibfnamefont{D.}~\bibnamefont{Caliste}},
  \bibinfo{author}{\bibfnamefont{R.}~\bibnamefont{Caracas}},
  \bibinfo{author}{\bibfnamefont{M.}~\bibnamefont{C\^ot\'e}},
  \bibnamefont{et~al.}, \bibinfo{journal}{Comput. Phys. Comm.}
  \textbf{\bibinfo{volume}{180}}, \bibinfo{pages}{2582 }
  (\bibinfo{year}{2009}).

\bibitem[{\citenamefont{Bottin et~al.}(2008)\citenamefont{Bottin, Leroux,
  Knyazev, and Z\'erah}}]{Bottin2008}
\bibinfo{author}{\bibfnamefont{F.}~\bibnamefont{Bottin}},
  \bibinfo{author}{\bibfnamefont{S.}~\bibnamefont{Leroux}},
  \bibinfo{author}{\bibfnamefont{A.}~\bibnamefont{Knyazev}}, \bibnamefont{and}
  \bibinfo{author}{\bibfnamefont{G.}~\bibnamefont{Z\'erah}},
  \bibinfo{journal}{Comput. Mat. Sci.} \textbf{\bibinfo{volume}{42}},
  \bibinfo{pages}{329 } (\bibinfo{year}{2008}).

\bibitem[{\citenamefont{Becke}(1988)}]{Becke1988}
\bibinfo{author}{\bibfnamefont{A.~D.} \bibnamefont{Becke}},
  \bibinfo{journal}{J. Chem. Phys.} \textbf{\bibinfo{volume}{88}},
  \bibinfo{pages}{1053} (\bibinfo{year}{1988}).

\bibitem[{\citenamefont{Perdew and Wang}(1992)}]{Perdew1992}
\bibinfo{author}{\bibfnamefont{J.~P.} \bibnamefont{Perdew}} \bibnamefont{and}
  \bibinfo{author}{\bibfnamefont{Y.}~\bibnamefont{Wang}},
  \bibinfo{journal}{Phys. Rev. B} \textbf{\bibinfo{volume}{45}},
  \bibinfo{pages}{13244} (\bibinfo{year}{1992}).

\bibitem[{\citenamefont{Torrent et~al.}(2008)\citenamefont{Torrent, Jollet,
  Bottin, Z\'erah, and Gonze}}]{Torrent2008}
\bibinfo{author}{\bibfnamefont{M.}~\bibnamefont{Torrent}},
  \bibinfo{author}{\bibfnamefont{F.}~\bibnamefont{Jollet}},
  \bibinfo{author}{\bibfnamefont{F.}~\bibnamefont{Bottin}},
  \bibinfo{author}{\bibfnamefont{G.}~\bibnamefont{Z\'erah}}, \bibnamefont{and}
  \bibinfo{author}{\bibfnamefont{X.}~\bibnamefont{Gonze}},
  \bibinfo{journal}{Comput. Mat. Sci.} \textbf{\bibinfo{volume}{42}},
  \bibinfo{pages}{337 } (\bibinfo{year}{2008}).

\bibitem[{\citenamefont{Dewaele et~al.}(2008)\citenamefont{Dewaele, Torrent,
  Loubeyre, and Mezouar}}]{Dewaele2008}
\bibinfo{author}{\bibfnamefont{A.}~\bibnamefont{Dewaele}},
  \bibinfo{author}{\bibfnamefont{M.}~\bibnamefont{Torrent}},
  \bibinfo{author}{\bibfnamefont{P.}~\bibnamefont{Loubeyre}}, \bibnamefont{and}
  \bibinfo{author}{\bibfnamefont{M.}~\bibnamefont{Mezouar}},
  \bibinfo{journal}{Phys. Rev. B} \textbf{\bibinfo{volume}{78}},
  \bibinfo{pages}{104102} (\bibinfo{year}{2008}).

\bibitem[{\citenamefont{Monkhorst and Pack}(1976)}]{Monkhorst1976}
\bibinfo{author}{\bibfnamefont{H.~J.} \bibnamefont{Monkhorst}}
  \bibnamefont{and} \bibinfo{author}{\bibfnamefont{J.~D.} \bibnamefont{Pack}},
  \bibinfo{journal}{Phys. Rev. B} \textbf{\bibinfo{volume}{13}},
  \bibinfo{pages}{5188} (\bibinfo{year}{1976}).

\bibitem[{\citenamefont{Nos\'{e}}(1984)}]{Nose1984}
\bibinfo{author}{\bibfnamefont{S.}~\bibnamefont{Nos\'{e}}},
  \bibinfo{journal}{J. Chem. Phys.} \textbf{\bibinfo{volume}{81}},
  \bibinfo{pages}{511} (\bibinfo{year}{1984}).

\bibitem[{\citenamefont{Gonze}(1997)}]{Gonze1997}
\bibinfo{author}{\bibfnamefont{X.}~\bibnamefont{Gonze}},
  \bibinfo{journal}{Phys. Rev. B} \textbf{\bibinfo{volume}{55}},
  \bibinfo{pages}{10337} (\bibinfo{year}{1997}).

\bibitem[{\citenamefont{Gonze and Lee}(1997)}]{Gonze1997a}
\bibinfo{author}{\bibfnamefont{X.}~\bibnamefont{Gonze}} \bibnamefont{and}
  \bibinfo{author}{\bibfnamefont{C.}~\bibnamefont{Lee}},
  \bibinfo{journal}{Phys. Rev. B} \textbf{\bibinfo{volume}{55}},
  \bibinfo{pages}{10355} (\bibinfo{year}{1997}).

\bibitem[{\citenamefont{Savrasov and Savrasov}(1996)}]{Savrasov1996}
\bibinfo{author}{\bibfnamefont{S.~Y.} \bibnamefont{Savrasov}} \bibnamefont{and}
  \bibinfo{author}{\bibfnamefont{D.~Y.} \bibnamefont{Savrasov}},
  \bibinfo{journal}{Phys. Rev. B} \textbf{\bibinfo{volume}{54}},
  \bibinfo{pages}{16487} (\bibinfo{year}{1996}).

\bibitem[{\citenamefont{Liu and Quong}(1996)}]{Liu1996}
\bibinfo{author}{\bibfnamefont{A.~Y.} \bibnamefont{Liu}} \bibnamefont{and}
  \bibinfo{author}{\bibfnamefont{A.~A.} \bibnamefont{Quong}},
  \bibinfo{journal}{Phys. Rev. B} \textbf{\bibinfo{volume}{53}},
  \bibinfo{pages}{R7575} (\bibinfo{year}{1996}).

\bibitem[{\citenamefont{Troullier and Martins}(1991)}]{Troullier1991}
\bibinfo{author}{\bibfnamefont{N.}~\bibnamefont{Troullier}} \bibnamefont{and}
  \bibinfo{author}{\bibfnamefont{J.~L.} \bibnamefont{Martins}},
  \bibinfo{journal}{Phys. Rev. B} \textbf{\bibinfo{volume}{43}},
  \bibinfo{pages}{1993} (\bibinfo{year}{1991}).

\bibitem[{\citenamefont{Jansen et~al.}(1977)\citenamefont{Jansen, Mueller, and
  Wyder}}]{Jansen1977}
\bibinfo{author}{\bibfnamefont{A.~G.~M.} \bibnamefont{Jansen}},
  \bibinfo{author}{\bibfnamefont{F.~M.} \bibnamefont{Mueller}},
  \bibnamefont{and} \bibinfo{author}{\bibfnamefont{P.}~\bibnamefont{Wyder}},
  \bibinfo{journal}{Phys. Rev. B} \textbf{\bibinfo{volume}{16}},
  \bibinfo{pages}{1325} (\bibinfo{year}{1977}).

\bibitem[{\citenamefont{Bauer et~al.}(1998)\citenamefont{Bauer, Schmid, Pavone,
  and Strauch}}]{Bauer1998}
\bibinfo{author}{\bibfnamefont{R.}~\bibnamefont{Bauer}},
  \bibinfo{author}{\bibfnamefont{A.}~\bibnamefont{Schmid}},
  \bibinfo{author}{\bibfnamefont{P.}~\bibnamefont{Pavone}}, \bibnamefont{and}
  \bibinfo{author}{\bibfnamefont{D.}~\bibnamefont{Strauch}},
  \bibinfo{journal}{Phys. Rev. B} \textbf{\bibinfo{volume}{57}},
  \bibinfo{pages}{11276} (\bibinfo{year}{1998}).

\bibitem[{\citenamefont{Johnson and Christy}(1972)}]{Johnson1972}
\bibinfo{author}{\bibfnamefont{P.~B.} \bibnamefont{Johnson}} \bibnamefont{and}
  \bibinfo{author}{\bibfnamefont{R.~W.} \bibnamefont{Christy}},
  \bibinfo{journal}{Phys. Rev. B} \textbf{\bibinfo{volume}{6}},
  \bibinfo{pages}{4370} (\bibinfo{year}{1972}).

\bibitem[{\citenamefont{Brorson et~al.}(1987)\citenamefont{Brorson, Fujimoto,
  and Ippen}}]{Brorson1987}
\bibinfo{author}{\bibfnamefont{S.~D.} \bibnamefont{Brorson}},
  \bibinfo{author}{\bibfnamefont{J.~G.} \bibnamefont{Fujimoto}},
  \bibnamefont{and} \bibinfo{author}{\bibfnamefont{E.~P.} \bibnamefont{Ippen}},
  \bibinfo{journal}{Phys. Rev. Lett.} \textbf{\bibinfo{volume}{59}},
  \bibinfo{pages}{1962} (\bibinfo{year}{1987}).

\bibitem[{\citenamefont{Mazevet et~al.}(2010)\citenamefont{Mazevet, Torrent,
  Recoules, and Jollet}}]{Mazevet2010}
\bibinfo{author}{\bibfnamefont{S.}~\bibnamefont{Mazevet}},
  \bibinfo{author}{\bibfnamefont{M.}~\bibnamefont{Torrent}},
  \bibinfo{author}{\bibfnamefont{V.}~\bibnamefont{Recoules}}, \bibnamefont{and}
  \bibinfo{author}{\bibfnamefont{F.}~\bibnamefont{Jollet}},
  \bibinfo{journal}{High Energy Density Physics} \textbf{\bibinfo{volume}{6}},
  \bibinfo{pages}{84 } (\bibinfo{year}{2010}).

\bibitem[{\citenamefont{Holst et~al.}(2011)\citenamefont{Holst, French, and
  Redmer}}]{Holst2011}
\bibinfo{author}{\bibfnamefont{B.}~\bibnamefont{Holst}},
  \bibinfo{author}{\bibfnamefont{M.}~\bibnamefont{French}}, \bibnamefont{and}
  \bibinfo{author}{\bibfnamefont{R.}~\bibnamefont{Redmer}},
  \bibinfo{journal}{Phys. Rev. B} \textbf{\bibinfo{volume}{83}},
  \bibinfo{pages}{235120} (\bibinfo{year}{2011}).

\bibitem[{\citenamefont{Pozzo et~al.}(2011)\citenamefont{Pozzo, Desjarlais, and
  Alf\`e}}]{Pozzo2011}
\bibinfo{author}{\bibfnamefont{M.}~\bibnamefont{Pozzo}},
  \bibinfo{author}{\bibfnamefont{M.~P.} \bibnamefont{Desjarlais}},
  \bibnamefont{and} \bibinfo{author}{\bibfnamefont{D.}~\bibnamefont{Alf\`e}},
  \bibinfo{journal}{Phys. Rev. B} \textbf{\bibinfo{volume}{84}},
  \bibinfo{pages}{054203} (\bibinfo{year}{2011}).

\bibitem[{\citenamefont{Rangel et~al.}(2012)\citenamefont{Rangel, Kecik,
  Trevisanutto, Rignanese, Van~Swygenhoven, and Olevano}}]{Rangel2012}
\bibinfo{author}{\bibfnamefont{T.}~\bibnamefont{Rangel}},
  \bibinfo{author}{\bibfnamefont{D.}~\bibnamefont{Kecik}},
  \bibinfo{author}{\bibfnamefont{P.~E.} \bibnamefont{Trevisanutto}},
  \bibinfo{author}{\bibfnamefont{G.-M.} \bibnamefont{Rignanese}},
  \bibinfo{author}{\bibfnamefont{H.}~\bibnamefont{Van~Swygenhoven}},
  \bibnamefont{and} \bibinfo{author}{\bibfnamefont{V.}~\bibnamefont{Olevano}},
  \bibinfo{journal}{Phys. Rev. B} \textbf{\bibinfo{volume}{86}},
  \bibinfo{pages}{125125} (\bibinfo{year}{2012}).

\bibitem[{\citenamefont{Lambert et~al.}(2011)\citenamefont{Lambert, Recoules,
  Decoster, Cl\'erouin, and Desjarlais}}]{Lambert2011}
\bibinfo{author}{\bibfnamefont{F.}~\bibnamefont{Lambert}},
  \bibinfo{author}{\bibfnamefont{V.}~\bibnamefont{Recoules}},
  \bibinfo{author}{\bibfnamefont{A.}~\bibnamefont{Decoster}},
  \bibinfo{author}{\bibfnamefont{J.}~\bibnamefont{Cl\'erouin}},
  \bibnamefont{and}
  \bibinfo{author}{\bibfnamefont{M.}~\bibnamefont{Desjarlais}},
  \bibinfo{journal}{Phys. Plasmas} \textbf{\bibinfo{volume}{18}},
  \bibinfo{eid}{056306} (\bibinfo{year}{2011}).

\bibitem[{\citenamefont{Faleev et~al.}(2006)\citenamefont{Faleev, van
  Schilfgaarde, Kotani, L\'eonard, and Desjarlais}}]{Faleev2006}
\bibinfo{author}{\bibfnamefont{S.~V.} \bibnamefont{Faleev}},
  \bibinfo{author}{\bibfnamefont{M.}~\bibnamefont{van Schilfgaarde}},
  \bibinfo{author}{\bibfnamefont{T.}~\bibnamefont{Kotani}},
  \bibinfo{author}{\bibfnamefont{F.}~\bibnamefont{L\'eonard}},
  \bibnamefont{and} \bibinfo{author}{\bibfnamefont{M.~P.}
  \bibnamefont{Desjarlais}}, \bibinfo{journal}{Phys. Rev. B}
  \textbf{\bibinfo{volume}{74}}, \bibinfo{pages}{033101}
  (\bibinfo{year}{2006}).

\bibitem[{\citenamefont{Chen et~al.}(2012)\citenamefont{Chen, Sametoglu, Tsui,
  Ao, and Ng}}]{Chen2012}
\bibinfo{author}{\bibfnamefont{Z.}~\bibnamefont{Chen}},
  \bibinfo{author}{\bibfnamefont{V.}~\bibnamefont{Sametoglu}},
  \bibinfo{author}{\bibfnamefont{Y.~Y.} \bibnamefont{Tsui}},
  \bibinfo{author}{\bibfnamefont{T.}~\bibnamefont{Ao}}, \bibnamefont{and}
  \bibinfo{author}{\bibfnamefont{A.}~\bibnamefont{Ng}}, \bibinfo{journal}{Phys.
  Rev. Lett.} \textbf{\bibinfo{volume}{108}}, \bibinfo{pages}{165001}
  (\bibinfo{year}{2012}).

\bibitem[{\citenamefont{Mueller and Rethfeld}(2013)}]{Mueller2013}
\bibinfo{author}{\bibfnamefont{B.~Y.} \bibnamefont{Mueller}} \bibnamefont{and}
  \bibinfo{author}{\bibfnamefont{B.}~\bibnamefont{Rethfeld}},
  \bibinfo{journal}{Phys. Rev. B} \textbf{\bibinfo{volume}{87}},
  \bibinfo{pages}{035139} (\bibinfo{year}{2013}).

\end{thebibliography}

\end{document}